\documentclass[useAMS]{emulateapj}
\usepackage{epsfig}
\usepackage{times}
\usepackage{natbib}
\usepackage{amsmath}

\newcommand{\kpc}{\>{\rm kpc}} 

\newcommand{\kms}{\>{\rm km}\,{\rm s}^{-1}} 
\newcommand{\Msun}{\>{\rm M_{\odot}}}
\newcommand{\hinv}{\>{\rm h}^{-1}}
\newcommand{\gyr}{\>{\rm Gyr}} 

\def\be{\begin{eqnarray}} \def\ee{\end{eqnarray}}
\def\ben{\begin{eqnarray*}} \def\een{\end{eqnarray*}}
\def\sec#1{Section~\ref{sec:#1}}
\def\fig#1{Figure~\ref{fig:#1}} 
\def\tab#1{Table~\ref{tab:#1}}
\def\equ#1{Equation~(\ref{equ:#1})}

\def\figp#1{Figure~\ref{fig:#1}}

\begin{document}

\title{On the origin of the  angular momentum properties of gas and dark
  matter in galactic halos and its implications}
\author{Sanjib Sharma, Matthias Steinmetz, Joss Bland-Hawthorn}
\affil{Sydney Institute for Astronomy, School of Physics, University of Sydney, NSW 2006, Australia}
\affil{Leibniz-Institut fuer Astrophysik Potsdam (AIP), An der Sternwarte 16, 14482, Potsdam, Germany}
\affil{Sydney Institute for Astronomy, School of Physics, University of Sydney, NSW 2006, Australia}

\begin{abstract}
We perform a set of non-radiative hydrodynamical simulations of
merging spherical halos in order to understand the angular momentum (AM)
properties of the galactic halos seen in cosmological simulations. 
The universal shape of AM distributions seen in
simulations is found to be generically produced as a
result of mergers. 
The universal shape is such that it has 
an excess of low AM material and hence 
cannot explain the exponential structure of disk galaxies. 
A resolution to this 
is suggested by the spatial distribution of low AM 
material which is found to be in the center and a conical region close 
to the axis of
rotation. A mechanism that preferentially
discards the material in the center and  prevents the material
along the poles from falling onto the disc is proposed as a solution. 
We implement a 
simple geometric criteria for selective removal of low 
AM material and show that in order for $90\%$ of halos to 
host exponential discs one has to reject at least $40\%$ 
of material.
Next, we explore the physical mechanisms responsible  for 
distributing the AM within the halo during 
a merger. For dark matter there is an 
inside-out transfer of AM, whereas for gas there is an outside-in
transfer, which is due to differences between collisionless 
and gas dynamics. This is responsible for the 
spin parameter $\lambda$ and the shape parameter $\alpha$ of AM
distributions being higher for gas as compared to dark matter.
We also explain the apparent high spin of dark matter halos 
undergoing mergers and show that a criteria stricter than what is currently 
used, would be required to detect such unrelaxed halos.  
Finally, we demonstrate that the misalignment of AM between 
gas and dark matter only occurs when the intrinsic spins of the
merging halos are not aligned with the orbital AM of the
system. The self-misalignment (orientation of AM when
measured in radial shells not being constant), which could be the 
cause of warps and anomalous rotation in disks galaxies, 
also occurs under similar conditions. 
The frequency and 
amplitude of this misalignment is roughly 
consistent with the properties of warps seen in disk galaxies.
\end{abstract}


\section{Introduction} \label{sec:intro2}
In the standard picture of galaxy formation,  
galactic halos acquire their angular
momentum (hereafter AM) via tidal torques \citep{1969ApJ...155..393P} in the linear regime and the process lasts till about
turnaround, when the system decouples from the Hubble flow. 
After the collapse the system forms a virialized 
structure. The gas inside the virialized dark matter halo then cools 
radiatively and collapses while conserving its AM, resulting 
in the  formation of centrifugally supported disks
\citep{1978MNRAS.183..341W,1980MNRAS.193..189F,1984ApJ...286...38W}. 
The process is also accompanied by the adiabatic contraction of the 
dark matter halo \citep{1986ApJ...301...27B}.
This standard picture leads to distribution of size and luminosity of galaxies 
in reasonable agreement with observations
\citep{1996MNRAS.281..475K,1997ApJ...482..659D,1998MNRAS.295..319M,1998ApJ...505...37A,2000ApJ...530..177V,2007ApJ...654...27D,2007ApJ...671.1115G}.

But detailed simulations revealed two problems. Firstly in simulations
incorporating gas with cooling and star formation, the gas was found to
lose a significant fraction of its AM, resulting 
in disks which were too small in size, a problem known as 
the angular momentum catastrophe
\citep{1991ApJ...380..320N,1994MNRAS.267..401N,1997ApJ...478...13N,1999ApJ...513..555S,1999ApJ...519..501S}. The
cause of the problem being that due to efficient cooling, the gas is
accreted as dense clumps which during mergers loses its AM
via dynamical friction.

Second problem is the angular momentum distribution (AMD hereafter)
problem, i.e., even if the AM is assumed to be conserved 
one cannot explain the exponential nature of disk galaxies.  
Using cold dark matter numerical simulations, it was shown by \citet{2001ApJ...555..240B}
that if disks are formed from gas with AMDs
similar to that of dark matter, then this results
in excess mass near the center as compared to an exponential disc.  
Specifically there is too much low
AM material and this makes it very hard to explain 
the origin of bulgeless dwarf galaxies \citep{2001MNRAS.326.1205V,2001MNRAS.327.1334V}. 
Simulations incorporation non-radiative gas also lead to 
similar conclusions \citep{2002ApJ...576...21V,2005ApJ...628...21S}. 
As demonstrated in \citet{2005ApJ...628...21S},  
the resulting AMDs written in terms of 
$s=j/j_{\rm tot}$ closely follows a law
of form $P(s)=[\alpha^{\alpha}/\Gamma(\alpha)]s^{\alpha-1}e^{-\alpha s}$, 
the universal form found in dark matter halos of cosmological $N$-body
simulations. Although the $\alpha$ parameter for gas is slightly
higher (close to $0.9$) than that of dark matter ($0.83$),    
it is still much less than  
what is needed for explaining the exponential structure of galactic
disks,i.e. , $\alpha > 1.3$.

The origin of the  
the universal form of the AMDs is still 
poorly understood and if we can understand it, that may provide the
clue to solving the problem.   
\citet{2002MNRAS.335..487M} proposed a model of build up of AM
by a sequence of mergers. In this model, the final 
halo spin is assumed to be the sum of orbital angular momenta of 
merging satellites.
The model was found to correctly reproduce the distribution of spin parameters
of halos \citep{2002ApJ...581..799V,2002MNRAS.329..423M}. A simple extension 
of this model was also found to roughly reproduce the AMDs. 
According to this model, 
the magnitude and direction of the total AM of 
a halo is predominantly determined by the  last major merger 
and hence the major merger contributes to the high AM
part of the AMD. The numerous small satellites 
fall in from random directions and mainly contribute to the low AM
  part of the AMD. 
This suggests that blowout of gas, e.g.,  by means of supernova feedback, 
from small halos can eliminate the low AM 
part of the distribution and may resolve the AMD 
problem in addition to the angular momentum catastrophe.

An alternative solution to the AMD problem
is that the 
feedback driven outflows preferentially discard low AM 
material during the assembly of the galaxy  \citep{2011MNRAS.tmp..595B}. 
In fact recent high resolution simulations including star formation and
feedback have been quite successful in forming bulge-less  exponential
disks \citep{2010Natur.463..203G,2011MNRAS.tmp..595B,2011arXiv1103.6030G} where such a
process has been shown to occur. Understanding the spatial
distribution of the low AM material may tell us 
as to which method is more effective in solving the AMD problem.

According to the model proposed by \citet{2002MNRAS.335..487M}, the most favorable 
scenario for galaxy formation is, where there are very few 
minor mergers, e.g., a halo acquiring its AM via a major merger. 
Is it enough to generate AMDs 
such that exponential disks can be formed? If the gas distribution 
is concentrated due to cooling or puffed up as with feedback, does 
it change the AMD of merger remnants? 
These are some of the questions that we investigate.

The AM properties of galaxies is of increasing interest
in observational surveys. For example, the Calar Alto Legacy Integral Field Area survey 
(CALIFA) will obtain spatially resolved spectroscopic information
 for about 600 galaxies in the local universe, using integral field spectroscopy
\citep{2011hsa6.conf..339S}. New imaging fiber bundles (so called
hexabundles) are to be used on wide-field survey telescopes
\citep[e.g., AAT;][]{2011OExpr..19.2649B,2011arXiv1112.3367C} to
obtain spatially resolved stellar and gas kinematics for a
volume-limited sample $10^{4-5}$ galaxies. With such surveys it will  
be possible 
to study the AMD of galaxies in voids, 
filaments, groups and clusters. 
Although current simulations which include star formation and feedback
have started showing success in forming 
disk galaxies, but these simulations are computationally very
expensive and this prohibits generation of a large sample of galaxies 
for statistical studies. On the other hand, dark matter only
simulations are computationally much less demanding which makes 
them suitable for comparison with large scale galaxy surveys, but 
one needs a way to populate dark matter halos 
with galaxies. Semi-analytic modelling of galaxies 
provides a way to do this
\citep{1994MNRAS.271..781C,1996MNRAS.283.1361B,1999MNRAS.303..188K,1999MNRAS.310.1087S,2000MNRAS.311..576K,2003ApJ...599...38B,2005ApJ...631...21K}.
However, most semi-analytic models do not take the low AM 
problem into account and this can have important consequences.
Additionally, it  is often 
assumed that the AM properties of gas is same as that 
of dark matter. This provides a motivation for studying  
the differences between the 
AM properties of gas and dark matter. 

The two main AM properties are the spin parameter 
and the shape of AMDs.
The spin of dark matter halos has been extensively studied 
and it has been shown that the distribution 
is well fit by a log normal distribution 
\citep{2002ApJ...576...21V,2005ApJ...628...21S,2007MNRAS.376..215B,2007MNRAS.381.1450N,2007MNRAS.378...55M,2010MNRAS.404.1137B}. 
In comparison there have been far fewer studies on AM 
properties of gas. 
In \citet{2005ApJ...628...21S} and \citet{2003ApJ...597...35C}, 
it was found that 
for halos simulated in cosmological context, 
the spin parameter and shape parameter of AMDs is higher for gas as
compared to that of dark matter.
Additionally, \citet{2005ApJ...628...21S}  found the spin ratio
$\lambda_{\rm Gas}/\lambda_{\rm DM}$ to increase
with cosmic time (at redshift zero the value being 1.4). 
The cause of this trend  
is still not known. In contrast, \citet{2002ApJ...576...21V} 
found the  
AM properties of gas and dark matter to be very
similar. A number of reason could be responsible for this
discrepancy. Firstly, it could be because 
\citet{2002ApJ...576...21V} in their analysis had 
included a large number of halos with low particle numbers. 
Secondly, 
\citet{2002ApJ...576...21V} had used thermally broadened gas
velocities to compare the AMDs of gas with that of dark matter,  
whereas broadening of velocities is known to mask out 
the differences between AMDs \citet{2005ApJ...628...21S}.  
Finally, \citet{2002ApJ...576...21V} had
analyzed the results at $z=3$ whereas the other authors had analyzed
them at $z=0$. 

Recently, it has been reported  that high spin halos
are more clustered than low spin halos \citep{2007MNRAS.376..215B,2010MNRAS.407..691D}. \citet{2007MNRAS.378...55M}
on the other hand do not find any environmental dependence. 
A crucial difference in the two schemes is the treatment of 
unrelaxed halos. It has been shown that  out-of-equilibrium halos 
tend to have higher spin and lower concentration, which 
when removed makes the halo concentration 
independent of spin \citep{2001ApJ...557..616G,2002ApJ...581..799V,2004MNRAS.348..921P,2006MNRAS.370.1905H,2007MNRAS.381.1450N}. 
Such an effect could also be responsible for
higher clustering of high spin halos. \citet{2007MNRAS.380L..58D} 
have studied the correlation of merger history and spin of halos 
and found that halos immediately after merging have higher spin. 
Later on during the virialization process the halos spin down due 
to redistribution of mass and AM.
Generally the offset of the
center of mass is used to parameterize the unrelaxed halos. 
How effective is this parameter in detecting unrelaxed halos? 
Observationally it is the spin of the baryonic component that is
observed, hence it is important to know if the gas also undergoes 
such a spin up and spin down during mergers?

Another area where gas shows a difference from dark matter is the issue 
of misalignment between them.
In non radiative hydrodynamical simulations the AM of gas in 
galactic halos is found to be 
misaligned with respect to dark matter with a mean angle of
$20^{\circ}$ \citep{2002ApJ...576...21V,2005ApJ...628...21S}.
In simulations with star formation and feedback the galactic discs 
are also found to be misaligned, with a median angle of
$\sim30^{\circ}$ \citep{2010MNRAS.404.1137B}.
The misalignment has important observational consequences.
For example it has been found that the distribution of satellite
galaxies is preferentially aligned along
the major axis of the central galaxy
\citep{2005ApJ...628L.101B,2006MNRAS.369.1293Y,2007MNRAS.376L..43A,2008MNRAS.385.1511W}. \citet{2006ApJ...650..550A}
show that if the disk AM vectors are aligned
with the minor axis of the halo or the AM of the halo
then the observed anisotropy can be
reproduced. \citet{2007MNRAS.378.1531K} further showed the second
option is preferred as orientation with minor axis results in a
stronger signal than that observed. If the AM of gas 
is misaligned with the dark matter then this could potentially lower 
the signal. 
Another example is related to the use of 
weak lensing studies to measure the projected mass density  
of a foreground galaxy in front of background galaxies. Since signal 
from an individual galaxy is weak, to produce detectable
signals, results of different galaxies are stacked together by 
orienting the images with respect to the shape of the central galaxy. If the
AM of the galaxies is misaligned with respect to the
shape of the dark matter halos, then this can wash out any ellipticity 
signal in the projected mass distributions \citep{2010MNRAS.404.1137B}. 

The AM vectors of gas and dark matter, 
in addition to being misaligned with each other, 
are also not perfectly aligned with
themselves within the halo
\citep{2005ApJ...627..647B,2010MNRAS.404.1137B}, 
which we refer to as self-misalignment.  
The self-misalignment is found to be most pronounced between the inner and
outer parts \citep{2005ApJ...627..647B}. 
For the gas such a self-misalignment could be
responsible for warps as seen in galactic discs.  
In recent cosmological hydrodynamical simulations,   
\citet{2010MNRAS.408..783R} show that the warps in their disks 
are due to the misalignment of the AM of the inner cold gas   
with that of the outer hot gaseous halo. 
Hence, it is important to understand as to when such a misalignment 
occurs.

The self-misalignment of AM
could also be responsible 
for the counter rotating gas as seen in some of the galaxies
\citep[][]{1995Natur.375..661C,2006AJ....131.1336S,2009ApJ...694.1550S}. 
Although recent mergers of gas rich systems are generally 
used to explain them, the models have some shortcomings. 
For example, if the merger is too massive it can heat up and thicken the disc
considerably; if it is small then in some cases 
it cannot account for all of the counter rotating gas 
\citep{1995Natur.375..661C,1996ApJ...461...55T}. 
Misaligned AM in galactic 
halos could provide an explanation for this.

To answer some of the questions posed earlier, 
we perform non-radiative hydrodynamical
simulations of merging 
spherical halos and analyze the AMDs 
 of the resulting remnant halos. 
We do simulations with various different 
orbital parameters and study the dependence of the shape 
parameter $\alpha$ of AMDs on these 
orbital parameters. We also analyze the ratio $\lambda_{\rm Gas}/\lambda_{\rm DM}$
and the misalignment angle $\theta$ of the remnant halos.

An outline of paper is as follows: in Section 2  
we  describe details of setting up initial conditions 
and methods of extracting AMDs
from halos; in Section 3 we investigate the AM 
properties of these halos. Finally, in Section 6 we 
summarize and discuss our results.

\section{Methods} \label{sec:methods}

 \begin{table*}
  \caption{ The initial set up parameters of merger simulations and their final properties.}
\footnotesize{
  \begin{tabular}{@{}|ccccccccccccccccc|}
\hline
\hline
Sim & $M_{\rm tot}$ & $f_m$     &   $\lambda_{\rm orb}$  &  $c_{\rm initial}$ &
  $T_{\rm orb}$ &   $\lambda_{\rm int}$ & $m_{\rm vir}$
  & $f^{\prime}_b$
  & $c_{\rm final}$ & $\theta$ & $\lambda_{\rm
      gas}$ & $\lambda_{\rm DM}$ &  $\lambda_{\rm
      gas}/\lambda_{\rm DM}$   & $\alpha_{\rm DM}$  
& $\alpha_{\rm Gas}$ & $\alpha_{\rm Gas}/\alpha_{\rm DM}$\\
\hline
1 & 100 & $0.5$ & $0.05$  & $10.0$ & $6.70$ & 0.0  & 83.6 & 1.0 & 10.4 &
0.4 & 0.037 & 0.039& 0.96(1.17)    & 0.85 & 0.97 & 1.14 \\ 
2 & 100 & $0.5$ & $0.05$  & $ 5.0$ & $6.70$ & 0.0  & 83.4 & 1.02 & 5.75 & 0.6
& 0.042 & 0.039& 1.08(1.27)    & 0.87 & 1.08 & 1.24 \\ 
3 & 100 & $0.5$ & $0.05$  & $15.0$ & $6.68$ & 0.0  & 84.4 & 1.0 & 15.1 & 0.8
& 0.035 & 0.038& 0.91(1.10)     & 0.78 & 0.90 & 1.15 \\ 
4 & 100 & $0.5$ & $0.01$  & $10.0$ & $6.40$ & 0.0  & 84.2 & 1.0 & 10.8 & 1.7
& 0.0079&0.0084& 0.94 (1.04)   & 0.89 & 0.93 & 1.04 \\ 
5 & 100 & $0.5$ & $0.10$  & $10.0$ & $7.50$ & 0.0  & 83.3 & 1.03 & 11.3 & 0.4
& 0.093 & 0.074& 1.25 (1.30)   & 0.86 & 1.02 & 1.19\\ 
6 & 100 & $0.5$ & $0.05$  & $10.0$ & $10.0$ & 0.0  & 80.2 & 1.02 & 10.5 & 0.1
& 0.043 & 0.037 & 1.15(1.49)    & 0.79 & 0.96 & 1.22 \\ 
7 & 100 & $0.1$ & $0.05$  & $10.0$ & $7.40$ & 0.0  & 91.8 & 0.95 & 9.2 & 2.0
& 0.046 & 0.024& 1.96(1.62)   & 0.82 & 0.74 & 0.90 \\ 
8 & 100 & $0.3$ & $0.05$  & $10.0$ & $6.60$ & 0.0  & 86.6 & 0.99 & 10.5 & 1.2
& 0.044 & 0.037& 1.17(1.28)    & 0.85 & 0.86 & 1.01 \\ 
9 & 167.2 & $0.5$ & $0.05$ & $10.08$ & $6.70$ & 0.039& 137.5 & 1.04 & 11.5 &
18.2 &0.044   & 0.041 & 1.09(1.22)   & 0.75 & 0.94 & 1.25 \\ 
10 & 100 & $0.5$ & $0.05$  & $ 10(1)$ & $6.70$ & 0.0  & 79.3 & 0.80 & 8.9 & 2.2
& 0.029 & 0.036& 0.79(0.68)    & 0.82 & 1.02 & 1.24 \\ 
11 & 100 & $0.5$ & $0.05$  & $10(25)$ & $6.70$ & 0.0  & 84.5 & 1.06 & 14.3 & 1.5
& 0.036 & 0.041 & 0.88(0.94)     & 0.81 & 0.91 & 1.12\\ 
\hline
\end{tabular}
}
\tablecomments{In the table, columns 2 to 7 are the parameters used 
to set up the initial conditions of the merger simulations. The columns are 
as follows: the total mass of the simulated
system $M_{\rm tot}=m_1+m_2$ ($10^{10} \hinv\Msun$), the fractional mass of the
least massive halo $f_{m}=\frac{{\rm min}(m_1,m_2)}{(m_1+m_2)}$, the initial spin
parameter of the whole system $\lambda_{\rm orb}$, 
the concentration parameter of the initial halos $c_{\rm initial}$, 
the orbital time period of the merging system $T_{\rm orb}$ ($\hinv\gyr$), 
and the intrinsic spin of merging halos $\lambda_{\rm int}$ .
Note, for Sim-10 and Sim-11 the  $c_{\rm initial}$ for gas is different from that of dark matter,
hence, the $c_{\rm initial}$ for gas is quoted in parenthesis. 
Columns 8 to 17 describe the properties of the merger remnants 
formed by the simulations. Here  $m_{\rm vir}$ ($10^{10} \hinv\Msun$)
is the virial mass of the remnant , $f^{\prime}_{b}$ is the 
baryon fraction within virial radius
relative to cosmological baryon fraction, $c_{\rm final}$ is the
concentration parameter of the remnant, $\theta$ (degree) the misalignment
angle between the dark matter and gas AM vectors. These are 
followed by the spin parameter $\lambda$ and the shape parameter 
$\alpha$ of AMDs for gas and dark matter. The quantity in brackets is the 
spin ratio at $t=6 \hinv\gyr$. 
The spin parameter is calculated using the definition 
given by \citet{2001ApJ...555..240B}.}
\label{tab:tb1}
\end{table*}

\subsection{Initial Conditions and Simulations} \label{sec:init_cond}
We study binary mergers of spherical halos consisting of dark matter 
and gas. The halos are set up with an exponentially truncated 
NFW density profile. 
\begin{eqnarray}
\rho(r) = \left\{
\begin{array}{ll}
\frac{\rho_s}{(r/r_s)(1+r/r_s)^2} & \textrm{for $r< r_{\rm tr}$} \\
\frac{\rho_s}{(r_{\rm tr}/r_s)(1+r_{\rm tr}/r_s)^2}\left(\frac{r}{r_{\rm tr}}\right)^\epsilon 
{\rm e}^{(-\frac{r-r_{\rm tr}}{r_d})} & \textrm{for $r> r_{\rm tr}$} 
\end{array}
\right. 
\end{eqnarray}
Imposing the condition that the logarithmic slope of $\rho$ at 
transition radius, $r=r_{\rm tr}$, should be continuous, gives 
\be
\epsilon= r/r_{\rm tr} -\frac{1+3c}{1+c}
\ee
For all our set ups, we use $r_d=0.1 r_{\rm vir}$. 
Normally, $r_{\rm tr}$ should be chosen to be equal to $r_{\rm rvir}$.  
However, the exponential truncation gives
rise to an extra mass. To compensate for this we choose 
$r_{\rm tr}$ to be less than $r_{\rm vir}$, such that the total
mass of the system is $m_{\rm vir}$.
For generating equilibrium realizations of the system, comprising 
of collisionless particles,  we 
follow the procedure given by \citet{2004ApJ...601...37K}.
In this procedure, first the phase space distribution function corresponding 
to a given density profile is numerically evaluated
and then the velocities of the collisionless particles 
are assigned by randomly sampling  this distribution. 
The gas is setup in hydrostatic equilibrium within the dark matter 
halo assuming a  density profile identical to that of the dark matter
($\rho_{\rm Gas}(r)= \rho_{\rm DM}(r) f_b/(1-f_b)$,
$f_b=\Omega_{\rm baryon}/\Omega_{\rm matter}$ 
being the cosmological baryon fraction). 
The thermal energy of the gas is given by
\be
u(r)=\frac{1}{\rho_{\rm Gas}(r)} \int_r^{\infty} \rho_{\rm Gas}(r)
\frac{G M(<r)}{r^2} dr,
\ee
$M(<r)$ being the cumulative mass enclosed within radius $r$.

In \tab{tb1} we list the parameters that are used to  set up
11 simulations, each with $N=2 \times 10^5$ dark matter particles and
an equal number of gas particles. Merger parameters were selected as 
follows.
A two body merger can be described in terms of the motion of a  
test particle with a reduced mass. A bound orbit of such a test
particle can be fully
characterized by semi major axis $a$ and eccentricity $e$ or 
equivalently by orbital time period $T_{\rm orb}$ and spin parameter 
$\lambda_{orb}$ of the system. The choice of these parameters is constrained 
by the fact that  $T_{\rm orb}$ should be less than the age of
universe and that the maximum separation between objects 
$r_{\rm rel}=a(1+e)$ should be greater than 
$r_{\rm vir1}+r_{\rm  vir2}$. The later condition provides a lower
bound on $T_{\rm orb}$. For our simulations, given a $\lambda$ 
we set $T_{\rm  orb}$ such that $r_{\rm rel}=r_{\rm vir1}+r_{\rm
  vir2}$. The exception being Sim-6 which has $r_{\rm rel}>r_{12}$ and 
hence was translated
analytically till the separation between the halos was equal to
$r_{12}$. Further details on the setup of merger parameters and its 
physical interpretation is given in \sec{merger_param} of the appendix.

For simulations 1 to 8 we assume the density distribution of gas to be same as
that of dark matter but in simulations 10 and 11 the gas is allowed to have 
a different density distribution, namely the concentration parameter
for gas is different from that of dark matter and this is shown in
brackets.
All the simulations except Sim-9
start with non-rotating halos, i.e., zero intrinsic spin. For the Sim-9 
we use the remnant halo obtained from Sim-1  as initial
halo and ${L_{\rm orb}}$ is set to be perpendicular to the 
spin ${L_{\rm int}}$ of the halos. 
The intrinsic halo spins are assumed to be
parallel to each other and are pointing towards the $z$ axis. 
For this setup  the direction of 
orbital AM in spherical coordinates 
is given by $(\phi,\theta)_{\rm  orb}=(-90,90)^{\circ}$. Three other
setups similar to this but with $(\phi,\theta)_{\rm
  orb}=(-90,45)^{\circ}$, $(-90,135)^{\circ}$ and $(-90,180)^{\circ}$
were also performed but are not listed in \tab{tb1}.

All simulations were evolved for $10 \hinv\gyr$.
The simulations were done using the smooth particle hydrodynamics 
code GADGET \citep{2001NewA....6...79S}. 
By construction, no assumptions on a particular background cosmology are made; 
however, for the NFW halo parameters we adopt the concordance 
$\Lambda$CDM cosmology 
with  $\Omega_{\lambda}=0.732$, $\Omega_{\rm m}=0.267$. A
gravitational softening of $2 \kpc \hinv$ was used.

In order to compare the AM properties of merger
simulations with those of simulations done in cosmological context we 
additionally use a set of 42 halos (virial masses between 
$1.3 \times 10^{11} M_{\odot}$ to  $1.5 \times 10^{13} M_{\odot}$),
which were selected from a $32.5 \hinv Mpc$ box length dark matter
simulation ($128^3$ particles), and were resimulated with gas 
at higher resolution by \citet{2005ApJ...628...21S} using GADGET. 
In these halos the number of dark matter particles 
within the virial radius ranges from 8000 to 80,000.
A gravitational softening of $2 \kpc \hinv$ was used.

\subsection{Calculation of Angular Momentum Distributions} \label{sec:methods1}
Dark matter particles 
are assumed to be collisionless and thus
a significant amount of random motions are superimposed onto the underlying 
rotational motion. So in order to calculate AMDs
, the velocity has to be smoothed 
 \citep[see][]{2005ApJ...628...21S}. 
Since the rotational motion is very small compared to the random
motion, one needs to smooth with a large number of neighbors. This large scale
smoothing introduces systematic biases which needs to be taken into
account. Smoothing the Cartesian components of velocity spuriously underestimates
the rotation for particles near the axis, as 
$ <v_{x}>= <v_{y}>=<v_{y}> \approx 0$ near the axis. To avoid this problem 
in \citet{2005ApJ...628...21S}  we smoothed the Cartesian components of 
AM vectors instead of velocities. 
As we will demonstrate later, the angular velocity 
$\Omega$ is nearly constant near the center. 
This implies that the AM vector $j$ has a strong, monotonically 
increasing radial dependence 
on cylindrical coordinate $r_c$. 
This results in an overestimate of the
AM of particles close to the axis. 
Existence of a strong radial density gradient further 
leads to underestimate of AM for particles along the equator.
To reduce some of these problems, in this paper we choose to smooth the
$z$ component of the angular velocity vector ${\bf \Omega_z}$ (the
halo being oriented such that the $z$ axis points along the total 
AM of the component being smoothed). 
A simple top hat kernel is
used for smoothing. The number of smoothing neighbors was chosen to be 
$5 \times 10^{-3}$ times the total number of particles within the virial
region. This makes the smoothing volume independent of the number of
particles in the halo. 
Note, smoothing is only employed to calculate the shape parameter $\alpha$
of the resulting AMDs.

\section{Angular Momentum Properties Of Remnant Halos} \label{sec:AMD}
The final properties of the merger remnants are given in \tab{tb1}.
We note that the virial mass $m_{\rm vir}$ of the remnant is less than the total
mass of the system $M_{\rm tot}$. Hence, a fraction of mass is lost.
The lost mass fraction is an increasing function 
of the kinetic energy $KE$ involved in the collision and a decreasing function of the 
total binding energy of the system. A detailed description of the mass
structure of remnant halos is given in \sec{mass_struct} in
appendix. In the present section we address the AM
properties of the remnant halos. First, we study the angular velocity 
and AMDs of the merger remnants and compare 
them with those of halos simulated in cosmological context. 
Next, we study the evolution of AM with time of our 
fiducial equal mass merger, Sim-1, and shed light on the physical 
mechanisms responsible for distributing the AM within
the halo. This is followed by studying the dependence of AM 
properties on orbital parameters. In the penultimate subsection we 
look into the issue of misalignment of AM vectors and in
the last subsection we study the spatial distribution of low 
AM material.

\subsection{Angular Velocity and Angular Momentum Distribution of
  Halos} \label{sec:angvel}

\begin{figure}
   \centering \includegraphics[width=0.5\textwidth]{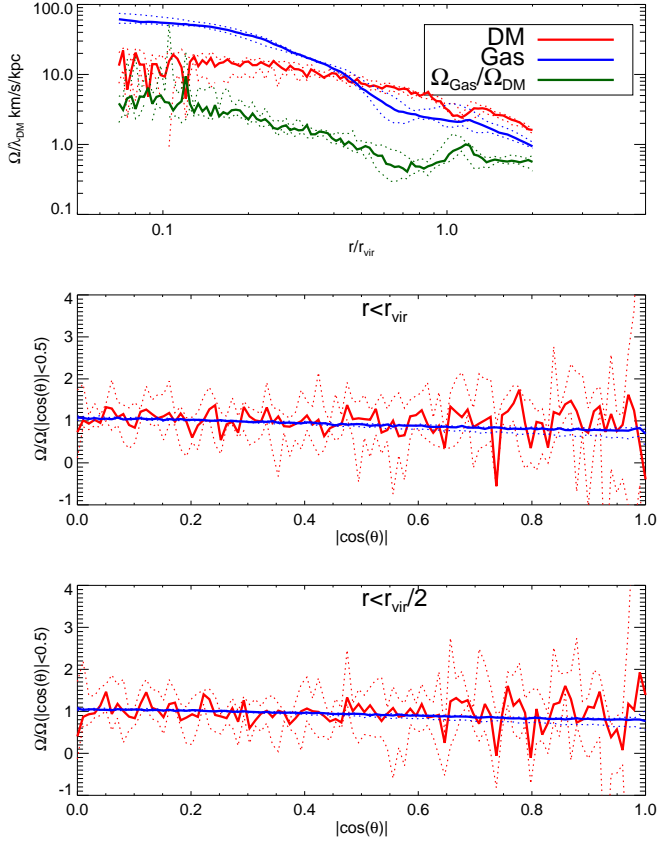}
   \caption{ Median angular velocity $\Omega=j_z/(x^2+y^2)$ as
     function of radius $r$ and angle $\theta$ for halos formed by
     mergers (merger simulations 1 to 8, excluding 4).  $\Omega$ as a function 
     of $r$ and $\theta$ was calculated by binning the particles so as
     to 1000 particles per bin. The
     dashed lines show $16$th and $84$th  percentile values.     
     The angular velocity
     profiles seems to flatten out for $r<0.2r/r_{\rm vir}$. In the top panel
     it can be seen  that in the inner regions the gas rotates
     faster than dark matter.
   Note, the gas profiles are much more smooth than that of dark matter 
   and this is because the 
   dark matter has significant amount of random
   motion superimposed on the actual rotation which is quite small.  
\label{fig:omega_r_theta_merger}}
\end{figure}

\begin{figure}
   \centering \includegraphics[width=0.5\textwidth]{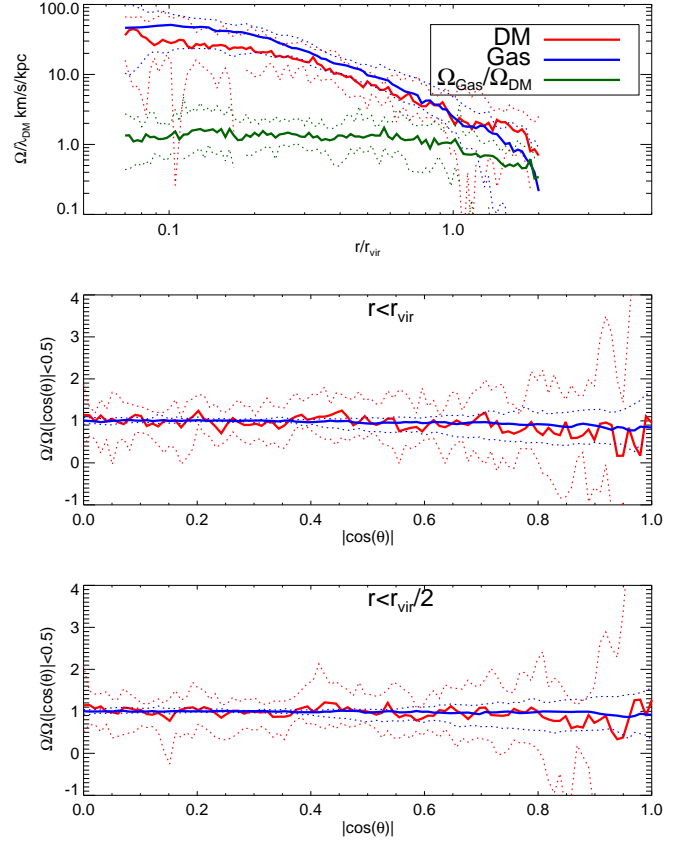}
   \caption{ Median angular velocity $\Omega=j_z/(x^2+y^2)$ as
     function of radius $r$ and angle $\theta$ for 21 halos simulated
     in a cosmological
     context. The halos where selected so as to have more than 30,000 
     particles within the virial radius individually for both gas and
     dark matter.  $\Omega$ as a function 
     of $r$ and $\theta$ was calculated by binning the particles so as
     to have 1000 particles per bin. The
     dashed lines show $16$th and $84$th  percentile values.     
     The angular velocity
     profiles seems to flatten out for $r<0.2r/r_{\rm vir}$. In the top panel
     it can be seen  that the gas rotates
     faster in the inner regions than that of dark matter. \label{fig:omega_r_theta_sim}}
\end{figure}

We explore the angular velocity $\Omega$ as a function of spherical coordinates
$r$ and $\theta$  for the remnant halos at $t=10
\hinv\gyr$, which represents the final
relaxed configuration. By angular velocity we mean 
the $z$ component of AM, with $z$ axis 
pointing along the direction of the total AM 
vector of the component being analyzed (gas or dark matter).
The angular velocity $\Omega$  of both gas and 
dark matter is found to be nearly independent 
of $\theta$, both for $r<r_{\rm vir}$ and  $r<r_{\rm vir}/2$ 
(lower two panels of \fig{omega_r_theta_merger}). 
The weak trend that exists is monotonic 
and favors faster rotation towards the equator.
This suggests 
that shells of matter are in solid body rotation. 
The top panel in \fig{omega_r_theta_merger} shows the  radial profiles of gas and dark matter.
In general $\Omega$ is a decreasing function of radius $r$ but the
profiles seem to flatten for $r<0.2r_{\rm vir}$.
As compared to dark matter, the gas is found to rotate faster in the
inner regions and slower in the outer regions.
For comparison the angular velocity profiles of  
halos simulated in cosmological simulations (from
\citet{2005ApJ...628...21S}) 
are shown in \fig{omega_r_theta_sim}. As in merger simulations they  
are nearly independent
of angle $\theta$, are a decreasing function of radius $r$, 
and show faster rotation for gas in the inner regions.  
However, the faster rotation for gas is not as strong as 
in merger simulations and the dip in gas rotation at 
about $r=0.7r_{\rm vir}$ is also not seen. Note, these are median
profiles: on a one to one basis the gas and
dark matter can show much more prominent differences as is revealed by
the fact that there is
significant scatter in the  ratio $\Omega_{\rm Gas}/\Omega_{\rm DM}$.  
Also a real halo has much more complex merger history which can
probably reduce the difference between dark matter and gas in the
inner regions. The difference in the outer parts is probably due to the
fact that the initial merging halos 
have an exponential
cut off in the outer parts whereas in real simulations the halos are much
more extended and moreover there is also smooth accretion onto the
halos. Hence, the outer parts of merger remnants may not be an accurate
representation of the real halos.

We now study the AMDs of the remnant halos at $t=10 \hinv\gyr$. 
The AM of each particle is obtained by smoothing its angular 
velocity with $400$ neighbors. For fitting the AMDs we use the following 
analytical function \citep[for details see][]{2005ApJ...628...21S} 
\be
P(j) & = &\frac{1}{j_{d}^{\alpha}\Gamma(\alpha)}(j)^{\alpha-1}e^{-j/j_{d}}
\textrm{where \ } j_d = j_{\rm tot}/\alpha. 
\ee
$j_{\rm tot}$ being the mean specific AM of the system.
Writing $P$ in terms of $s=j/j_{\rm tot}$ and replacing $j_{d}$ the
cumulative distribution reads as 
\begin{eqnarray}
  P(<s) & = & \gamma(\alpha,\alpha s)
\label{equ:gamma}
\end{eqnarray}
where $\gamma$ is the Incomplete Gamma function.
In \citet{2005ApJ...628...21S} this function was used to fit the AMDs 
of halos obtained in cosmological simulations and of model 
exponential disks embedded in NFW potentials. 
It was found that for AMDs of exponential disks embedded in NFW potentials 
the shape parameter $\alpha$ 
is greater than $1.3$ whereas for cosmological halos values are typically
smaller than $1$ ($<\alpha_{\rm DM}>=0.83$ and $<\alpha_{\rm
  gas}>=0.89$ ). For dark matter for fiducial Sim-1 we find 
$\alpha=0.85$ whereas for others it is given by $0.75<\alpha<0.9$. 
For gas in Sim-1, $\alpha$ is 0.97 and for others it is
between $0.74$ and $1.08$.  The gas has significantly 
larger $\alpha$ than dark matter and this is because of the fact that 
the gas rotates faster than dark matter in the inner regions.
Merger simulations successfully reproduce the fact that 
$\alpha_{\rm Gas}>\alpha_{\rm DM}$ as in cosmological simulations.
If we take Sim-1 as the fiducial case then for dark matter the 
value of $\alpha$ is in excellent agreement with cosmological
simulations 
but for gas we find that it is about $8\%$ higher.  
As discussed earlier the gas in merger simulations are an idealized
case and in real halos the gas rotation profiles are slightly flatter 
in the outer parts and this explains the slightly lower $\alpha$ in them.

\subsection{Evolution of Angular Momentum with Time} \label{sec:jevol}
\begin{figure}
    \centering \includegraphics[width=0.5\textwidth]{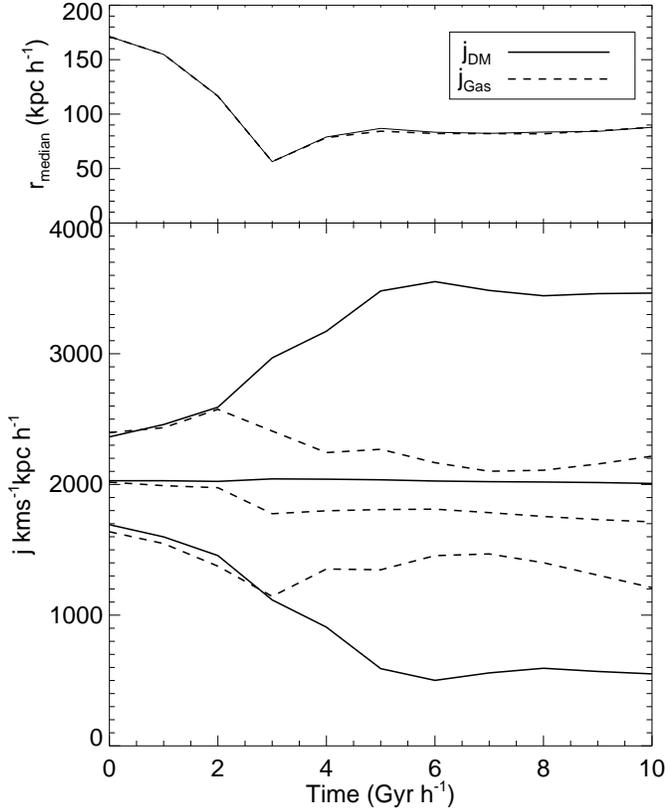}
   \caption
{ 
 The top curve shows the evolution of the half mass radius
 $r_{\rm median}$. The bottom three curves show the evolution of the specific
 angular momentum $j$ of the gas and dark matter components for Sim-1 (an equal mass merger) for
 the full system and separated into the inner half mass and outer half mass.
\label{fig:j_vs_t_m50}}
\end{figure}

\begin{figure*}
    \centering \includegraphics[width=0.9\textwidth]{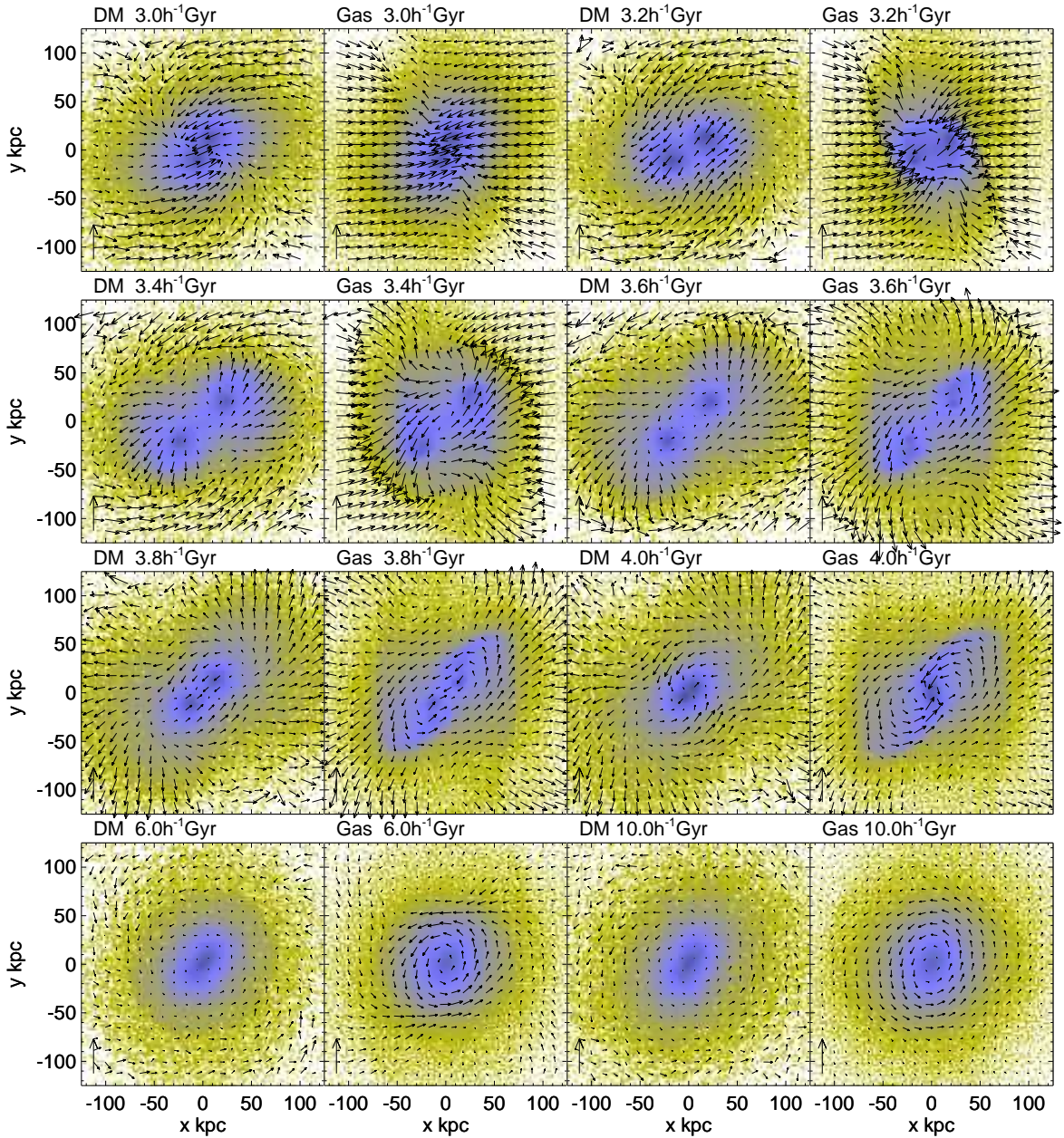}
   \caption{ Velocity vector field in the X-Y plane along with density
     map for gas and dark
     matter at various stages during the evolution of an equal mass merger
     simulation (Sim-1). Field is shown for particles within a radius of 125 kpc and
     $|z|<20$. The maximum length of the arrow corresponds to $250 \kms$
     and is shown in left hand corner of each panel.  At 3.2 and 3.4 
      $\hinv\gyr$ it can be seen that at the regions
     near the shocks (top right and lower left) the in-falling gas 
     in the outer parts is transferring angular momentum to the inner
     parts. By $t=$ 6 $\hinv\gyr$ the gas can be seen to 
     rotate faster in the inner regions as compared  to dark
     matter. 
\label{fig:vel_field}}
\end{figure*}

In this section we analyze the time evolution of the specific AM  of gas 
and dark matter components for Sim-1 (\fig{j_vs_t_m50}), 
which is an equal mass merger of halos with concentration
parameter of $10$, $T_{\rm orb}=6.7 \hinv\gyr$ and $\lambda_{\rm
  initial}=0.05$. We use this as a fiducial case to understand the
main properties of the evolution, specifically the origin of the
differences in the AM  properties of gas and dark matter.
We analyze  the evolution of the  total AM as well as that
of the  inner and the outer parts of equal mass, separated by median radial
distance. 
The evolution of median radius $r_{\rm median}$, separating
the two halves is also plotted alongside. 
We divide the evolution into four stages, Stage 1 from $0-2 \hinv\gyr$, 
Stage 2 from $2-3 \hinv\gyr$, Stage 3 from $3-6 \hinv\gyr$ and 
Stage 4 from $6-10 \hinv\gyr$.

During a merger the system first collapses to a compact configuration
(Stage 1  and Stage 2), marked by  
a decrease in $r_{\rm median}$ (see also \fig{vel_field}). In Stage 3
the system expands,  as shown by slight increase in $r_{\rm median}$,
and then in Stage 4
the system evolves without 
any significant change in the density structure.
It can be seen from the evolution of the total AM of the  system
(middle lines in \fig{j_vs_t_m50}) that in Stage 2 the gas 
loses about $10\%$ of
its AM to dark matter. The dark matter gains AM in this stage but its change is quite
small since the mass of dark matter is much larger than that of gas.
In Stage 3 and 4 the total AM of gas and dark matter is nearly constant.  
In contrast, when the AM of inner and outer half masses are  analyzed 
separately, significant differences can be seen.
For dark matter there is an inside-out transfer of AM in stages 1, 2 and 3.
Due to dynamical friction the inner part loses AM continuously to the outer
part until it virializes to form a pseudo equilibrium distribution
after which the
evolution stops. It can be seen that the AM evolution of gas is
decoupled from dark matter, from stage $2$ 
onwards. Initially,  both the inner and outer parts of gas lose AM
to dark matter. However, in Stage 3, when the inner parts start to expand, for the gas the AM is
transferred to the inner parts from the outer parts. The fact that the
rise in AM of gas in inner parts is almost the same as the fall in AM of
gas in outer parts means that the transfer of AM is purely between the gas 
components. This transfer is because the expanding inner part of gas 
that also has low AM, shocks with in-falling outer part that has high AM, 
thereby leading to transfer of AM. This is visible more clearly in 
\fig{vel_field} where we plot the velocity
field in the X-Y plane within a radius of $125 \kpc$ and  $|z|<20$.
At 3.0 $\hinv\gyr$ the halos can be seen crossing each other and 
at 3.2 $\hinv\gyr$ they have 
crossed and are now pushing against the outer
material of the other halo which is still falling in. The outer material 
falling in from upper right and lower left corners 
pushes and transfers AM to the expanding inner regions. With time the
shocks progressively move outwards. 

In contrast the dark matter cannot
shock, their particles can cross each other 
and they exchange energy and AM via violent relaxation. 
It is easier to understand their evolution 
in terms of  an inside-out spherical collapse
simulation in which the inner regions collapse faster than the outer
regions.  
In such a system as described in \citet{2008gady.book.....B} a high 
energy particle in
outer region falls into a gradually 
steepening potential and hence gains kinetic energy. Later when it starts to 
move out the inner region has already expanded and hence it has to 
climb out of a shallower potential. The net result of all this is that 
a high energy particle that falls in late gains energy. 
Now, the impact parameter of the particle during collision
is high for particles in the outskirts that are falling in late.   
Since the AM of a particle is proportional to the 
impact parameter it is also high for them.
Hence, high 
AM particles mostly end up orbiting in the outer regions. 
In contrast for gas these
late falling
high AM material shocks and transfers its AM to the inner parts.

Overall the conclusion is that due to gas dynamical effects the baryons are
more efficient in depositing the AM to the inner parts of the halo 
making it rotate faster than dark matter in the inner regions.
The faster rotation of the gas in the inner region is 
visible in \fig{omega_r_theta_merger}) and also 
in the bottom panels of \fig{vel_field}. 
The faster rotation of gas also responsible for 
$\alpha_{\rm Gas}$ being greater than $\alpha_{\rm DM}$.

The above hypothesis suggests that increasing the energy of the
collision or the 
orbital AM should 
make the outside-in transfer of AM for gas and inside-out transfer of 
AM for dark matter more stronger. The Sim-6, which is same as Sim-1
except for the fact that it has higher $T_{\rm orb}$ meaning more
energetic merger, does reveal this. As expected, the ratios of 
 $\lambda_{\rm Gas}/\lambda_{\rm DM}$ and
 $\alpha_{\rm Gas}/\alpha_{\rm DM}$ are found to be higher for Sim-6. 
Similarly, when considering 
Sim-4, Sim-1 and Sim-5  having lowest intermediate and  
highest orbital AM one finds that the 
ratios of  $\lambda_{\rm Gas}/\lambda_{\rm DM}$ and
 $\alpha_{\rm Gas}/\alpha_{\rm DM}$ increases monotonically (see \tab{tb1}).
These results thus provide support to the above hypothesis.

In the final Stage 4 of the evolution  
the system has almost reached a pseudo equilibrium. 
During this stage, for the gas there is  a gradual transfer of AM from
the fast rotating inner layers to the slow rotating outer layers.

\begin{figure}
   \centering \includegraphics[width=0.5\textwidth]{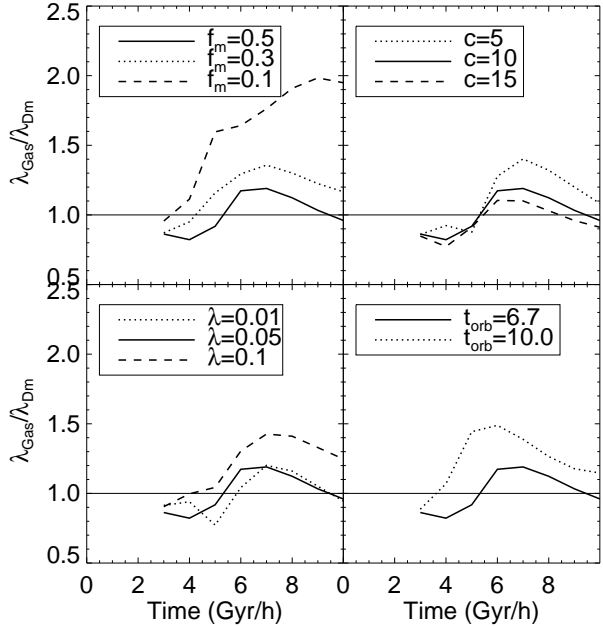}
   \caption[Evolution of $\lambda_{\rm Gas}/\lambda_{\rm DM}$ with time for 
     various merging scenarios.]{ Evolution of 
$\lambda_{\rm Gas}/\lambda_{\rm DM}$ with time for 
     various merging scenarios.
\label{fig:fj_t}}
\end{figure}

\subsection{Dependence of Spin Ratio $\lambda_{\rm Gas}/\lambda_{\rm
    DM}$ on Orbital Parameters and its Evolution with Time} \label{sec:lambda_ratio}
Having understood the AM evolution in the inner and outer parts, we
now try to understand the evolution of AM within the virial radius, 
which is commonly employed to measure the spin of the halos.  
\fig{fj_t} describes the evolution of the spin ratio
$\lambda_{\rm Gas}/\lambda_{\rm DM}$ with time for various different
merging scenarios. Note, the spin parameter is calculated using the
definition $\lambda={  j_{\rm tot}}/\sqrt{2 G m_{\rm vir} r_{\rm vir}}$
\citep{2001ApJ...555..240B}, where ${j_{\rm tot}}$ is the specific
AM of the material within virial radius $r_{\rm vir}$, 
and $m_{\rm vir}$ is the virial mass.
At each stage of the evolution we identify the virial region by means of 
the spherical over-density criterion and then compute the relevant 
properties of the virialized remnant halo.
In Stage 1 ($0-2 \hinv\gyr$) of the evolution
the ratio is close to $1$.  In Stage 2 ($2-3 \hinv\gyr$) the ratio drops by
about $10-20\%$. 
In Stage 3 ($3-6 \hinv\gyr$),  the ratio rises and reaches a peak
at around $6-7 \hinv\gyr$ and then in Stage 4 the ratio decreases (except for Sim-7).  
It is easy to understand the time evolution of spin ratios in the
context of the discussion done
earlier in \sec{jevol}.
In Stage 1 the gas and dark matter have not yet decoupled so the ratio 
is close to $1$.  In Stage 2 the gas  loses its AM
to dark matter and hence a drop in the spin ratio.
In Stage 3, in the inner regions the  
dark matter loses AM while the gas gains, this results in a 
rise in the spin ratio.  
Finally, in Stage 4 the AM  of dark matter in the inner regions remains nearly
constant whereas for gas there is an inside-out transfer and this again results 
in a drop in the spin ratio.

Next, we study the dependence of spin ratio on the orbital parameters.
In each of the panels in \fig{fj_t}, we vary one of the orbital parameters 
(namely $f_m,c,\lambda$ and $T_{\rm orb}$) while keeping the other parameters 
identical to that of benchmark Sim-1.
In top left panel we compare Sim-1, Sim-8 and Sim-7,
having $f_m=$0.5, 0.3 and 0.1 respectively, $f_m=m_2/(m_1+m_2)$
being the mass fraction of the smaller merging halo.
At a given time the gas to dark matter spin ratio is found to
be higher for a  lower value of $f_{m}$. For $f_m=0.1$ it continues to
increase  even in Stage 4 and reaches a value as high as 2. 

In the second panel, i.e., top right, we plot the results for mergers with 
different values of concentration parameter, Sim-1, Sim-2 and Sim-3
having $c_{\rm initial}=$10.0, 5.0 and 15.0 respectively. 
Lower concentrations yield higher
spin ratios. It can be seen from \tab{tb1} that $\lambda_{\rm DM}$ is
largely unaffected by the change in $c_{\rm initial}$ whereas
$\lambda_{\rm Gas}$ increases with lowering the concentration.  
In Sim-10 and Sim-11 we vary the concentration parameter of gas,  
setting it to 1 and 25 respectively, and keep 
the concentration of dark matter constant at 10.0. The Sim-10  
is designed to mimic the case of a halo where the gas is puffed up by feedback 
from star formation whereas Sim-11 mimics the case where the gas has 
cooled and collapsed to the central regions. 
\tab{tb1} shows that 
when considering the total AM content, 
the concentrated gas loses more AM than the puffed
gas. This demonstrates the AM catastrophe problem in
which due to excessive cooling the gas gets concentrated 
and during subsequent evolution lose AM as a result 
of dynamical friction. Surprisingly, when AM is measured with
in the virial region the puffed up gas has less AM. 
This is because for the puffed up case significant amount of gas is
outside the virial radius and this gas also has high AM 
whereas for the concentrated  gas case all the gas ends up within
the virial radius. This is reflected in the baryon fraction as shown
in \tab{tb1} which is 0.8 for the former and 1.06 for the later.

In the bottom left panel 
we look at the role of varying the orbital AM. 
We compare Sim-1, Sim-4 and Sim-5 having a  $\lambda_{\rm orb}=$0.05, 0.01
and 0.10 respectively. Increasing $\lambda_{\rm orb}$ beyond $0.05$ increases the
spin ratio while lowering it does not affect the results significantly.
Finally, we investigate the role played by the kinetic energy
associated with the collision, 
which is controlled by varying the parameter $T_{\rm orb}$. Larger $T_{\rm orb}$ 
means that halos approach each other from a farther distance and have more
energetic collision.  We compare Sim-1 and Sim-6 which have $T_{\rm orb}=$ 6.7 and 10.0 respectively.  
More energetic collision leads to higher spin ratio. 
In light of discussion in \sec{jevol} the effects discussed above 
are due to the fact that late infalling 
gas, in case of high $T_{\rm orb}$ and $\lambda_{orb}$,
 shocks  more strongly leading to more transfer of AM to inner parts 
and for dark matter the late infalling gas is more energetic and is more likely
to escape outside the virial radius.

In general it can be seen that at around $6 \hinv\gyr$ i.e., $3 \hinv\gyr$
after the merger the ratio $\lambda_{\rm Gas}/\lambda_{\rm DM}>1$ for
all merging scenarios and this provides an explanation for the results of
\citet{2005ApJ...628...21S} and \citet{2003ApJ...597...35C} where they find
$\lambda_{\rm Gas}/\lambda_{\rm DM}\sim 1.4$ and $1.2$ respectively 
for halos simulated in a cosmological context.

\begin{figure}
   \centering \includegraphics[width=0.5\textwidth]{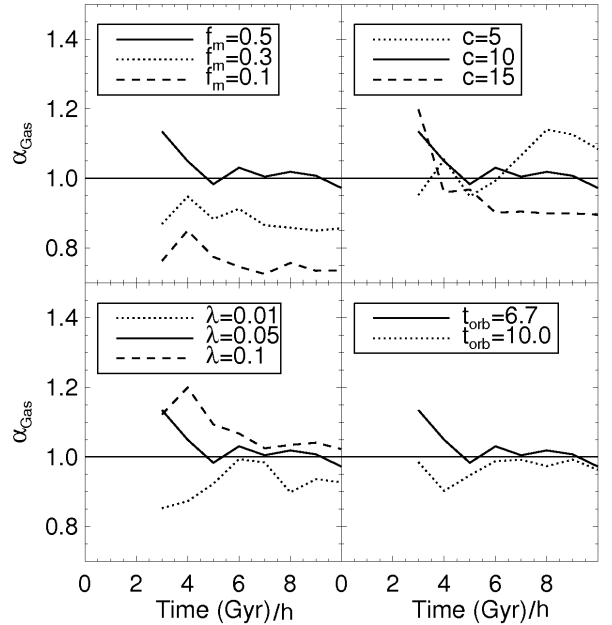}
   \caption{ Evolution of $\alpha_{\rm Gas}$ with time for 
     various merging scenarios. $\alpha$ is calculated from the 
     AMDs obtained from 
     smoothing the motion of particles obtained from simulations. 
\label{fig:al_t}}
\end{figure}

\subsection{Dependence of Shape Parameter  $\alpha$ on Orbital Parameters}
In this section we explore the role of the orbital parameters on the shape
parameter $\alpha$ of gas obtained by fitting the AMD of the remnant
halos by \equ{gamma}.
In \fig{al_t} we plot the evolution of the shape parameter $\alpha_{\rm Gas}$ for various merging scenarios.
The comparisons done in various panels are the same as  in \fig{fj_t}.
The values of $\alpha_{\rm Gas}$ below $t=3 \hinv\gyr$ are not relevant for the study here
since the merger has not yet happened. 
Between $3-5 \hinv\gyr$ there is a slight variation where the halo is still
relaxing, but beyond that  for all orbital geometries  $\alpha$
 has very little evolution with time (\fig{al_t}). 
As an apparent trend, $\alpha$ decreases slightly with time ( except Sim-2).
Varying the parameter  $\lambda_{\rm orb}$ or $t_{\rm orb}$ does not seem to affect
the values of $\alpha$. 
Decreasing the mass ratio $f_m$ decreases the value of $\alpha$, while
decreasing the concentration parameter $c_{\rm initial}$ increases its
value. Varying only the concentration of gas as in Sim-10 and Sim-11 
also has similar effect (see \tab{tb1}), namely puffed up halos have 
higher $\alpha$ whereas concentrated halos have lower $\alpha$.  
In the context of the AMD problem  this 
means puffing up gas by means of feedback can partially 
help to resolve the problem, but the value of $\alpha=1.02$ is still 
far short of that required to form exponential disks ($\alpha>1.3$). 
Hence, just by itself the puffing up of gas 
is not enough to solve the problem.

Finally, we note that the gas in general has higher $\alpha$ than that of dark
matter and the  $\alpha_{\rm Gas}/\alpha_{\rm DM}$
increases with increase of $T_{\rm orb}$ and $\lambda_{\rm orb}$  
(see \tab{tb1}).  As discussed  in  \sec{jevol} this is due to the 
differences between 
gas dynamics and collisionless dynamics. 
The fact that  $\alpha_{\rm Gas}$ is greater than $\alpha_{\rm DM}$, 
is consistent with the findings of
\citet{2003ApJ...597...35C} and \citet{2005ApJ...628...21S} for cosmological halos, and our results here provide an explanation for it.

\begin{figure*}
   \centering \includegraphics[width=0.75\textwidth]{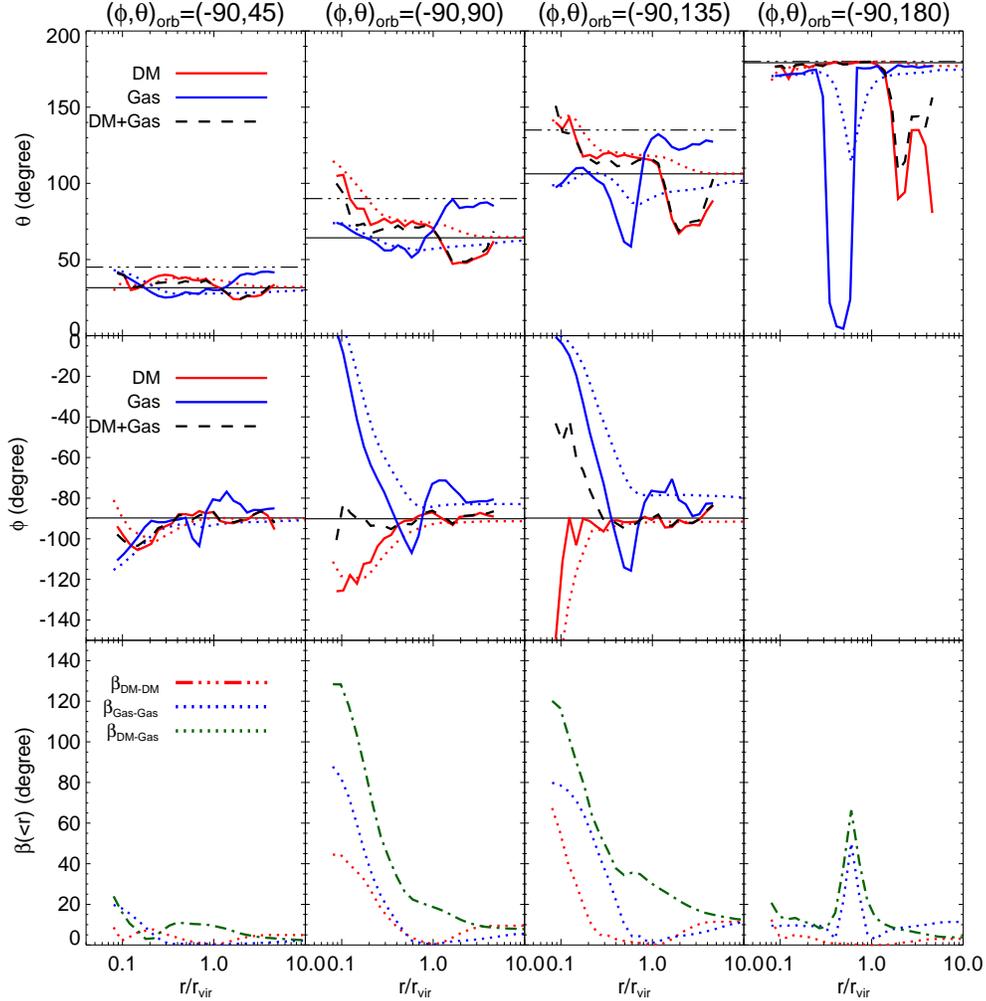}
   \caption{ Orientation angles $\theta$ and $\phi$ of angular
     momentum vectors (of gas and dark matter) for different merging 
     scenarios. In each case the intrinsic spin of merging halos 
     points along the $z$ axis. Four cases with different directions 
     of orbital angular momentum are shown.   
     In the upper two panels, the horizontal black solid line shows the direction of
     the total angular momentum of the system while the dot dashed
     line shows the direction of orbital angular momentum. The curves were
     obtained by binning the particles  in 25  logarithmically spaced
     bins in radial distance. The dotted lines show the cumulative profiles while the
     solid lines are differential profiles. The dashed line is the
     differential profile of the whole system (gas and dark matter). 
     The bottom row shows the cumulative misalignment angle of dark 
     matter with itself, gas with itself and between gas and dark matter. 
     The gas and dark matter angular momentum vectors are
     significantly misaligned with each other, especially in the inner
     regions. Additionally, in the inner region $r<0.2r_{\rm vir}$, the angular
     momentum vectors of gas and dark matter are also individually 
     misaligned with their total angular momentum vectors.
 \label{fig:misalignment}}
\end{figure*}

\subsection{Misalignment of the angular momentum vectors of gas and dark matter} 
The misalignment angle $\theta$ for all  simulations is tabulated in
\tab{tb1}. Mergers with zero intrinsic spins do 
not seem to generate any significant misalignment in the final 
remnant halo. The misalignment angle $\theta$ is less than $2^{\circ}$ for 
all orbital geometries.
 Now, if we imagine a merger of
 halos with non-zero intrinsic spins,  then the final angular 
momentum can be though of as  ${\bf j}={\bf j}_{\rm
  int}+{\bf j}_{\rm orb}$, $j_{\rm int}$ being the intrinsic specific 
AM and $j_{\rm  orb}$ the orbital specific AM.
Due to differences between 
gas and collisionless dynamics we already know $j_{\rm
  orb}^{\rm Gas}>j_{\rm orb}^{\rm DM}$, similarly the contribution 
of the intrinsic components could also differ. 
If ${\bf j}_{\rm int}$ is aligned
with ${\bf j}_{\rm orb}$ then there is no possibility to generate any
misalignment. However, if they are not aligned then we can 
expect to see misalignment, except for the special case where 
$ j_{\rm orb}^{\rm Gas}/j_{\rm orb}^{\rm DM}=j_{\rm int}^{\rm Gas}/j_{\rm
  int}^{\rm DM}$.

To test the above scenario, in Sim-9 
we merge two halos (extracted from Sim-1) having non-zero  
spin and an orbital AM which is perpendicular to
the spin. 
The remnant halo is found to be significantly
misaligned with a misalignment angle close to $17^{\circ}$. In 
\fig{misalignment} we show the orientation of the AM
vectors of gas and dark matter as measured in radial shells 
for various different directions of the
orbital AM. 
We mainly concentrate on regions with $r>0.1r_{\rm vir}$ which should
be quite reliable given that our gravitation softening is about 
$0.01r_{\rm vir}$.
The intrinsic spin has the direction
$\theta=0.0$. The solid and dashed lines are the differential profiles
while the rest are cumulative profiles. In the top panel the lower
horizontal line marks the mean expected $\theta$  for the halo assuming uniform
mixing. The upper line shows the angle for the orbital AM.
The gas and dark matter show very different trends. For dark matter the inner 
region is dominated by orbital  AM, the outer by intrinsic spin
and the middle region has intermediate direction. For the gas the inner
region has intermediate values, the outer region is dominated by
orbital AM and the middle region is dominated by
intrinsic spin, which points towards $\theta=0^{\circ}$. In the 
rightmost column corresponding to a retrograde merger 
the gas even shows a spin flip in the middle regions.

In the bottom panels it can be seen that the cumulative misalignment
angle defined as $\beta_{\rm
  DM-Gas}(<r)={\rm cos}^{-1}(\hat{\bf j}_{\rm DM}(<r).\hat{\bf j}_{\rm DM}(<r))$
increases inwards into the halo. The cumulative self-misalignment
of the AM 
, $\beta_{\rm DM-DM}$ and  $\beta_{\rm Gas-Gas}$,  which is measured
with respect to the AM within the virial radius also shows
similar trend. In the bottom row the magnitude of misalignment
increases from left to right, i.e., with increase of angle between orbital and
intrinsic AM.

In the panels in second column, the case of $\theta_{\rm orb}=90$,
 it can be seen 
that most of the misalignment is due to the $\phi$ direction 
varying sharply in the inner regions.
Moreover, the gas and and dark matter AM vectors seem
to be pointing in opposite directions in $\phi$. 
This is surprising given the expected value of $\phi$ is $-90^{\circ}$. 
It is not clear if
the gas is being torqued by dark matter or is it simply rearrangement
of AM. To understand this cumulative profiles are 
plotted as dotted lines. By $r=r_{\rm vir}$ the $\phi$ seems to have 
averaged to the expected value for both gas and dark matter.
Also, in $\theta$ the cumulative profiles tend to the expected value
at large $r$ for
both dark matter and gas. This suggests redistribution and self
torquing to be the main mechanism for the variation of the gas AM direction.
However, beyond $r_{\rm vir}$, $\phi$ for gas is slightly larger 
than 90, hence some amount of torque must have been exerted on it from dark matter. 
The panels in other columns also lead to similar conclusion.

\begin{figure}
   \centering \includegraphics[width=0.5\textwidth]{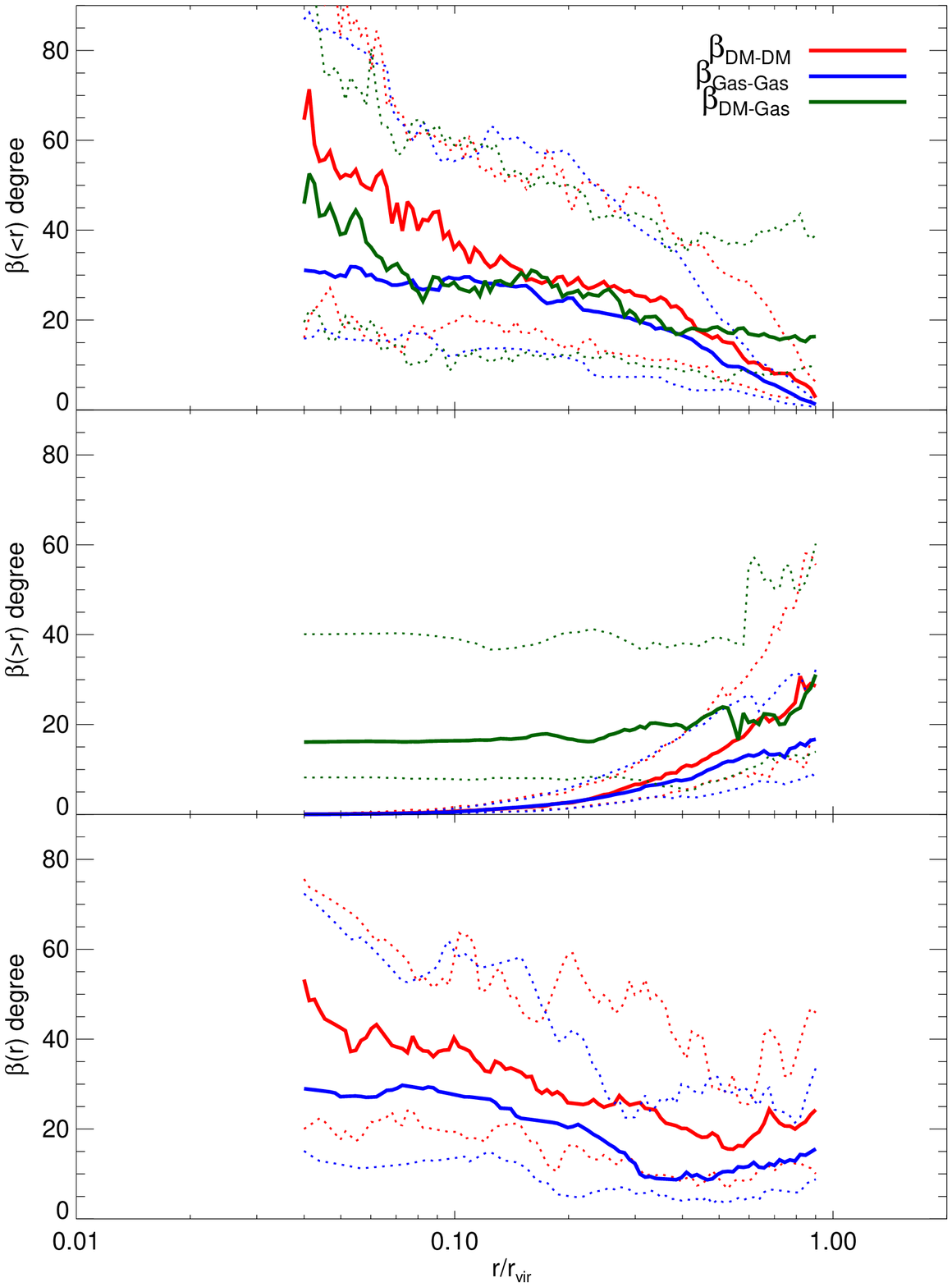}
   \caption{Angular momentum orientation profiles of gas and dark
     matter for halos simulated  in cosmological context. The top 
     panel shows the cumulative profile with
     $\beta(<r)=\hat{\bf \rm j}(<r).\hat{\bf \rm j}_{\rm vir}$, the middle also shows
     the cumulative profile but for $\beta(>r)=\hat{\bf \rm
       j}(>r).\hat{\bf \rm j}_{\rm
       vir}$  and the lower panel shows the differential profile with
     $\beta(r)=\hat{\bf \rm j}(r).\hat{\bf \rm j}_{\rm
       vir}$. The angle between angular momentum vectors of gas and
     dark matter is defined as $\beta_{\rm DM-gas}(<r)=\hat{\bf \rm
       j}_{\rm DM}(<r).\hat{\bf \rm j}_{\rm Gas}(<r)$. $\beta_{\rm
       DM-gas}(>r)$ also having an analogous definition. 
 \label{fig:misalignment_sim}}
\end{figure}

We now compare the results of cosmological simulations with that 
of merger simulations.
In \fig{misalignment_sim}, in the top panel we plot the 
cumulative misalignment angles $\beta_{\rm DM-DM}(<r),\beta_{\rm
  Gas-Gas}(<r)$ and $\beta_{\rm DM-Gas}(<r)$ as defined earlier. 
The trends in the top panel are similar to the trends in the second column
of bottom row in \fig{misalignment} which corresponds to the most
probable orientation of a merger. 
\fig{misalignment_sim} shows that the median misalignment of gas with
respect to dark matter is about $20^{\circ}$ which is 
reproduced by Sim-9. Note, higher misalignments can also be achieved 
if $\theta_{\rm orb}$ is greater than $90^{\circ}$. 
The fact that results of the  cosmological simulations 
are successfully reproduced by the merger simulations leads us to
conclude   that 
the difference in gas and 
collisionless dynamics is the main cause of misalignment of
AM vectors as seen in cosmological simulations.
Moreover, misalignments occur only when the intrinsic spins are not 
aligned with the orbital AM.

Top panel in \fig{misalignment_sim} also shows that the misalignments
are more pronounced in the inner parts  than the outer parts. 
The fact that the total AM vectors are dominated 
by the AM in the outer parts is partly responsible for 
this.  Finally, in the orientation profiles of cosmological
simulations the 
dark matter shows more misalignment than the gas whereas the opposite 
was true for the merger simulations. Given that a real halo has a much 
more complex merger history than that of a single merger as shown here, we
do not consider the discrepancy to be too significant.  
Moreover there is a  significant scatter about the median profiles 
as shown in \fig{misalignment_sim}, 
which means  that on a one to one basis the gas and dark matter can have
different trends suggesting that they are sensitive to the merger
history and hence could be employed to understand them.

We now compare the amount of misalignment seen in our
cosmological halos with those of previous studies. 
The previous studies were mostly confined to dark matter only 
simulations hence only $\beta_{\rm DM-DM}$ can be compared.
\citet{2010MNRAS.404.1137B} show the cumulative misalignment 
angle with respect to AM vector of material 
within $r<0.25r_{\rm vir}$ while \citet{2005ApJ...627..647B} use a
differential distribution.
In order to facilitate comparison with earlier studies we 
additionally plot in middle and bottom panels 
the cumulative $\beta(>r)$ profiles and the differential $\beta(r)$ 
profiles. 
For the range of masses considered here 
\citet{2010MNRAS.404.1137B} find a value of around
$\beta_{\rm DM-DM}(r<0.25r_{\rm vir})=25^{\circ}$ (their
Fig 4). The top panel of our figure shows the corresponding quantity to 
be  $30^{\circ}$ which is in good agreement with their results.
\citet{2005ApJ...627..647B} plot the orientation profiles with respect to
AM measured in different shells. 
If the shell at 
$r=0.6r_{\rm vir}$ is  taken be the representative of the total AM, 
 then this gives a value of 25 and 35 for $\beta_{\rm
  DM-DM}(0.1r_{\rm vir})$ and $\beta_{\rm
  DM-DM}(r_{\rm vir})$  which is again in
very good agreement with profile shown in bottom panel of our 
\fig{misalignment_sim}. 

Our result for misalignment between gas and dark matter, median
$\beta_{\rm DM-Gas}(<{\rm r_{vir}})=20^{\circ}$, are in agreement with those of 
\citet{2002ApJ...576...21V}. This was also reported in
\citet{2005ApJ...628...21S}. These results are for non-radiative 
gas.
A portion of this gas would cool and later on form 
disc galaxies. Additional, physical processes like star formation 
and feedback could further alter the angular momentum of the disc. 
Hence, in general we expect real galaxies to be even more misaligned. 
In fact, \citet{2010MNRAS.404.1137B} using 
simulations with star formation and feedback , find 
the median misalignment between the AM of the central
galaxies and the DM halo to be $\sim30^{\circ}$.

\subsection{Spatial Distribution of Low and Negative Angular Momentum Material}
\begin{figure*}
   \centering \includegraphics[width=0.9\textwidth]{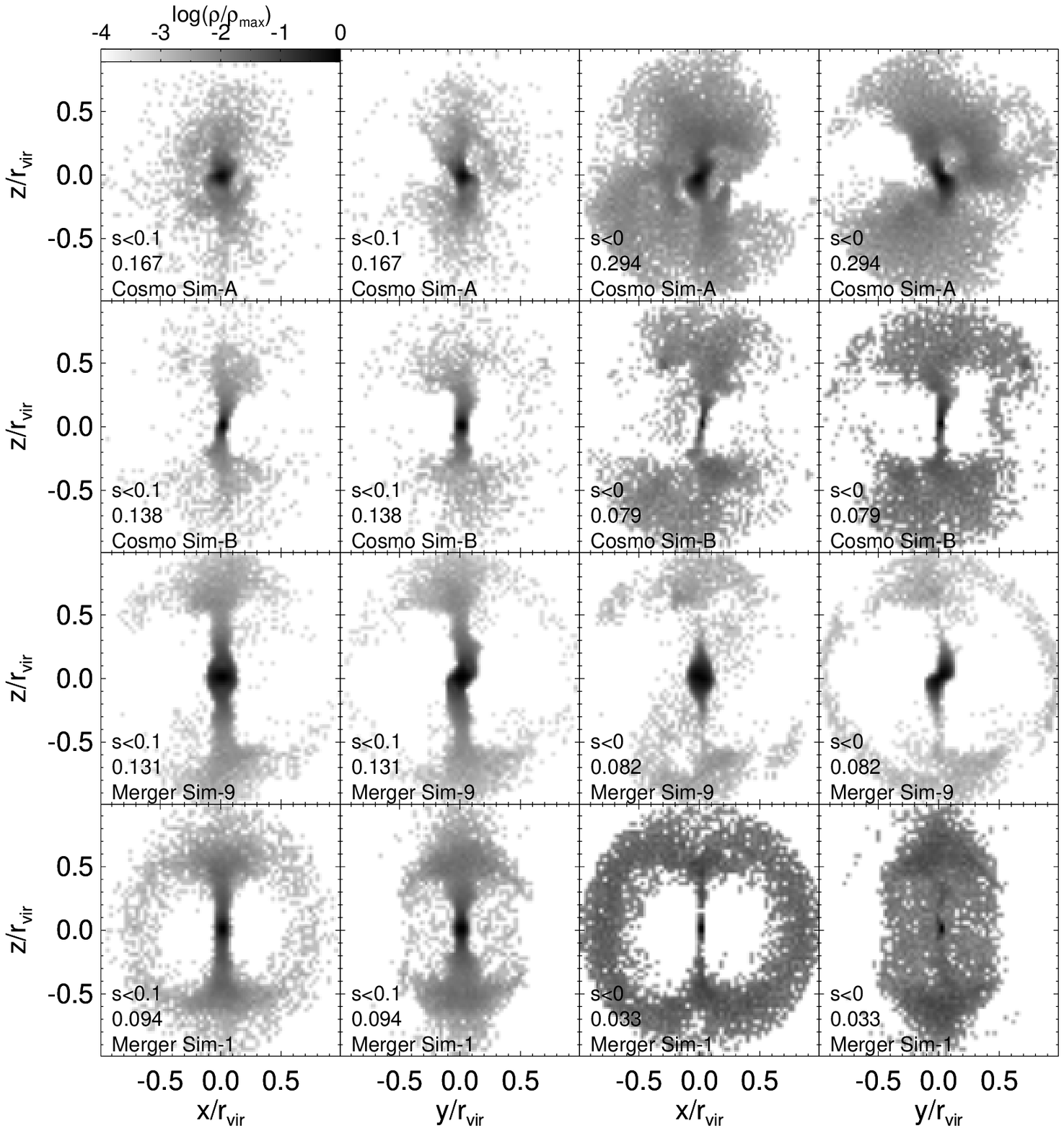}
   \caption{
       Spatial density maps of particles with low AM, i.e., $0<s<0.1$
       (columns one and two)  and negative AM, i.e.,$s<0$) (columns
       three and four) in various halos within the virial radius.
     Particles are shown in $x-z$ and $y-z$ plane with $z$ axis pointing in the
     direction of angular momentum. The fraction
     of low and negative AM particles is also labelled on each plot.
     The grey scale showing the density maps is
     normalized to the maximum density in each plot. 
     The top two panels are for halos from cosmological
     simulations while the lower two panels are for halos formed by
     merger simulations, namely, Sim-9(third row) and Sim-1 (fourth row). 
     The merger simulations results are shown for the final relaxed
     configuration at $t=10\hinv\gyr$.
     Particles with low and negative angular momentum are concentrated
     in the center and along the axis of rotation. 
\label{fig:xyz_low_neg_bw}}
\end{figure*}

\begin{figure}
   \centering \includegraphics[width=0.5\textwidth]{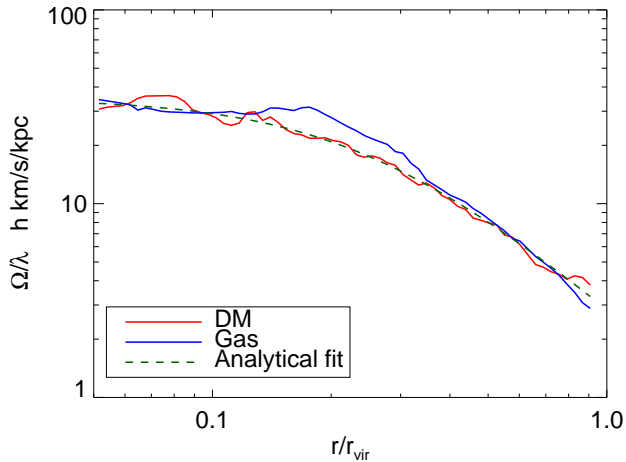}
   \caption{ The angular velocity $\Omega$ as a function of radial 
distance for both gas and dark matter. 
The dashed line is the analytical expression that 
provides a good approximation to the profiles.
\label{fig:omegafit}}
\end{figure}

\begin{figure}
   \centering \includegraphics[width=0.5\textwidth]{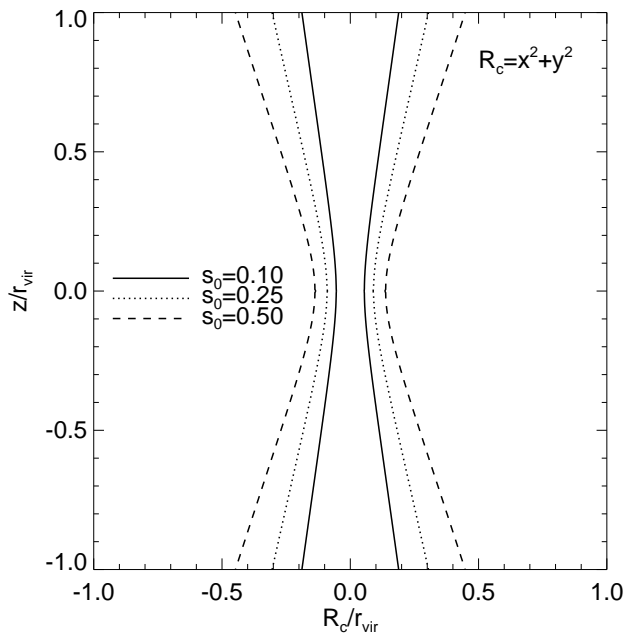}
   \caption{ The shape in $(R_c,z)$ plane of the volume removed to get 
rid of the low AM material in galactic halos. The shape is controlled 
by the parameter $s_0$ which is the minimum angular momentum,  
as given by \equ{amd_exp}, that is retained. 
\label{fig:remove_geometry}}
\end{figure}

\begin{figure}
   \centering \includegraphics[width=0.5\textwidth]{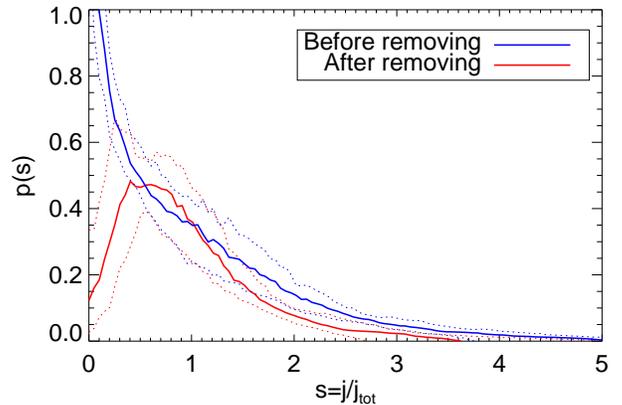}
   \caption{The effect of selectively removing low angular momentum 
material on the AMDs. The material was removed using \equ{amd_exp} 
with $s_0=0.5$. The solid curves show the mean AMDs computed over 
42 halos and the dotted curves show the 16 and 84 percentile values. 
Note, $j_{\rm tot}$ is not same for both the cases.
In fact $j_{\rm tot}$ increases after removal of low AM material.
Also, for the after removal case,  
the area under the curve is less than one, the area 
actually represents the fraction of mass retained. 
\label{fig:removal_effect_amd}}
\end{figure}

\begin{figure}
   \centering \includegraphics[width=0.475\textwidth]{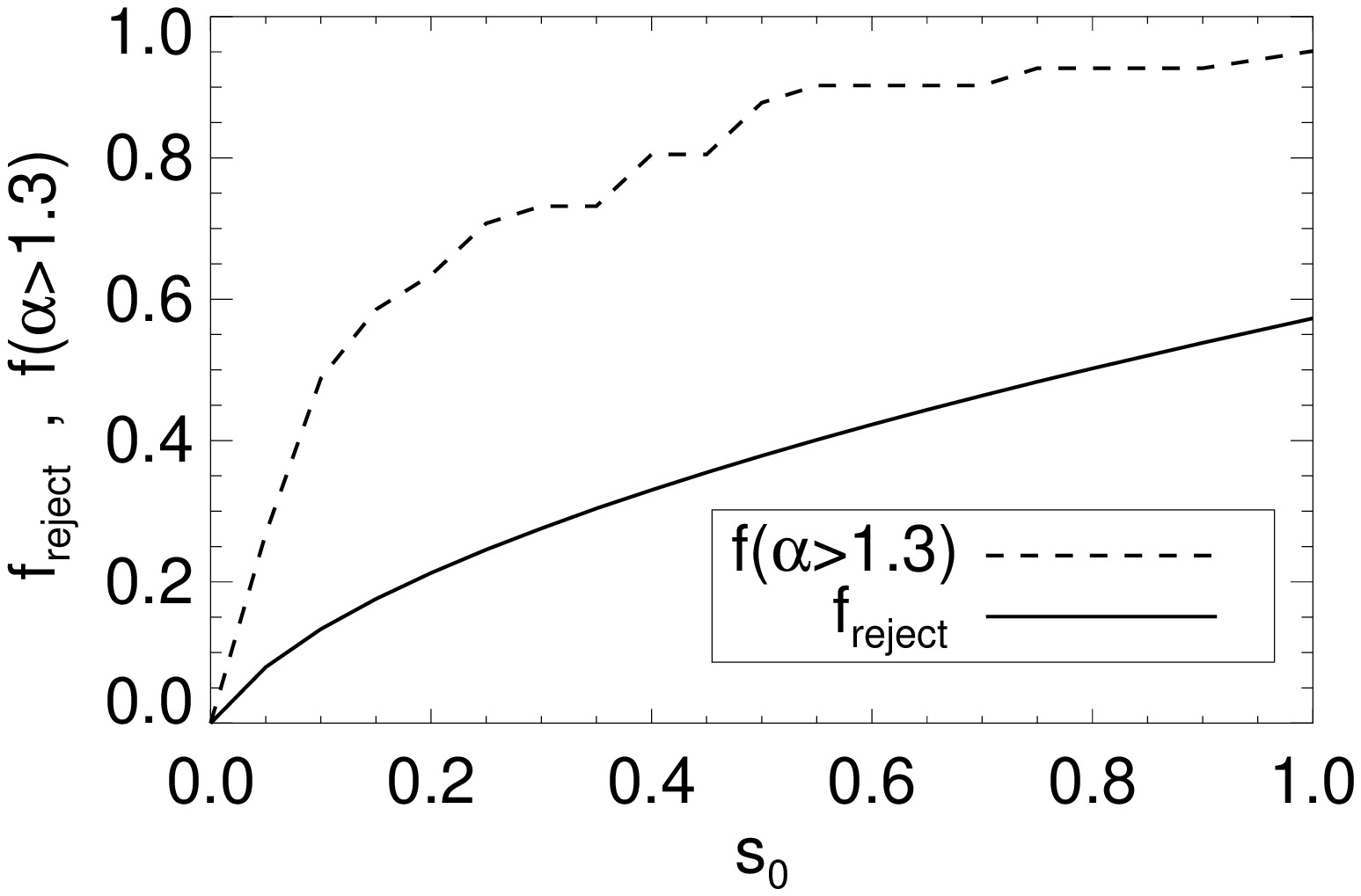}
   \caption{The variation of fraction of lost mass, $f_{\rm reject}$, 
and the fraction of halos having $\alpha>1.3$ with the parameter 
$s_0$ which controls the amount of material removed.  
\label{fig:removal_effect}}
\end{figure}

The AMD of gas in halos simulated in a cosmological context, shows an excess of
low AM material as compared to the AMD required to form
an exponential disc. If $s=j/j_{\rm tot}$ is the specific AM normalized to
the mean specific AM, then $0<s<0.1$  
is the typical region where the theoretical prediction differs from 
that of the exponential disc. Hence, we select particles in this range and
study their distribution in space.
In addition, there is also the issue of 
material with negative AM, which can arise from two
sources. The first source is random turbulent motions and the second
source is 
large scale flows that are remnants of shocks and misaligned 
minor mergers occurring after the major merger. 
The negative AM due to
the former source would be typically in regions with low AM 
and 
would vanish when velocities are smoothed locally, as is done while 
calculating the AMDs. On the other hand, large scale flows cannot be
easily smoothed and is the main reason why
\citet{2005ApJ...628...21S} find that, in spite of smoothing,   
cosmological halos have about $8\%$ of matter in negative AM. 
During the assembly of the disk the negative AM is going to further
enhance the fraction of low AM material, hence it is also important to 
study its distribution in the current context.

In \fig{xyz_low_neg_bw}  we show the x-z and
y-z density maps of low and negative AM gas particles as defined above
in various halos. AM is computed from raw un-smoothed velocities.
In the plots the $z$ axis is aligned with the total AM 
vector of the halo.
The top two rows are for halos from cosmological simulations  
whereas the lower two rows  are for merger simulations. Among these, the 
third row is for a merger where intrinsic spins are misaligned with the 
orbital AM (Sim-9), and the fourth row is for the fiducial case of an
equal mass zero intrinsic spin merger (Sim-1).   
The plots show that the low AM material 
is near the center and  in a conical region around the rotation
axis. The negative AM material is also in regions where the AM is low.
As discussed earlier in \sec{angvel}, angular velocity $\Omega$ is nearly independent of angle 
$\theta$ and is a decreasing function of radial distance $r$. 
In terms of cylindrical radius $R_{\rm c}=\sqrt{x^2+y^2}$ the AM 
is given by $\Omega R_{\rm c}^2$. Hence, the
AM is low along the axis
of rotation. The conical shape is due to the fact that $\Omega$ is
a decreasing function of $r$. 
Assuming negative AM is due to 
random turbulent motions, one expects it to be in regions where
$\Omega R/v_{\rm random}$ is small, which would again be similar to
the distribution of low AM material. 

For the remnant halo in Sim-1, a part of the negative AM material
of gas is distributed in a ring shaped structure in the x-z plane.
During a merger, a plane of compressed and  shocked gas is formed which 
is ejected out radially. The ring is created when such a gas which has
very low AM falls back at a later time. Note, only about
$3\%$ of the gas is in such form which after smoothing reduces to $1\%$.

The spatial distribution of low AM material has a dependence 
on the merger history of the halos.
For example, in the third row, which is a merger of halos with 
intrinsic spins misaligned with the orbital AM, the central region
looks more puffed up in x-z projection in comparison to the halo in the
fourth row. In y-z projection one can 
see the that the central region is twisted. This appearance is
because the AM in the inner regions is misaligned with respect
to the total AM. The cosmological halo in the top row also shows such a 
behavior suggesting a major merger with misaligned spins. 
The halo in the second row also has slightly twisted axes in the inner
region but is very similar to the halo in the bottom row suggesting 
that the intrinsic spins of its progenitors were either small or
well aligned with the orbital AM.

The characteristic distribution of low AM material found in remnant halos as
well as cosmological halos suggests that during galaxy formation a mechanism 
which preferential ejects material from the central regions and
prevents further material from collapsing along the
rotation axis may alleviate the AMD
 problem. Such preferential ejection may be
possible with feedback from star formation or AGN. Essentially, the inner
parts would collapse first and start forming stars. The feedback would
then drive a radial outflow, but since the assembling gas 
will have a flattened configuration with density being highest near the 
equatorial regions, the outflow would 
naturally be stronger along the poles causing preferential ejection of
low AM material.

To further investigate this idea, we devise a simple geometric 
criteria to selectively remove the low
AM material 
and then check the resulting AMDs to see if they conform to those 
of exponential discs, i.e., $\alpha>1.3$. 
Firstly, we fit the  
the angular velocity as a function of radial distance for both 
gas and dark matter by a simple analytical expression.
The profiles were found to be well approximated by 
the following equation (see \fig{omegafit}).
\be
\frac{\Omega_z}{\lambda^\prime} &=&\frac{f_0}{1+(r/r_0)^{1.75}} \ \
 {\rm h^{-1}} \kms/\kpc
\ee
where $f_0=35$, $r_0=0.25 r_{\rm vir}$ and 
\be
\lambda^\prime &=& \frac{j_{\rm tot}}{j_{\rm vir}}=\frac{j_{\rm tot}}{\sqrt{\Delta_{\rm vir}} H r_{\rm vir}^2}
\ee
Here, $\Delta_{\rm vir}=18\pi^2+82(\Omega_{\rm m}-1)-39(\Omega_{\rm m}-1)^2$ is
the virial over-density parameter used to calculate the virial region for a given 
matter density $\Omega_{\rm m}$ of the universe
\citep{1998ApJ...495...80B}, and $H=0.1 \kms{\rm
  kpc}^{-1}{\rm h} $ is the hubble constant.
Note, $j_{\rm tot}$ is the mean specific AM of each component.
For $\Omega_m=0.267$ 
the expected specific AM is then given by
\be
s' &=&\frac{j_z}{j_{\rm tot}}=35.61\frac{(R_{\rm c}/r_{\rm
    vir})^2}{1+(r/r_0)^{1.75}} 
\label{equ:amd_exp}
\ee
where $R_{\rm c}=\sqrt{x^2+y^2}$ is the cylindrical radius.
Particles with $s'>s_0$ 
are selected for removal. The resulting geometry  
for various values of $s_0$  is shown in \fig{remove_geometry}. 
The geometry is conical in shape and resembles those seen in outflows. 
For a fiducial setting of $s_0=0.5$, the AMDs
before and after removal are shown in
\fig{removal_effect_amd}. It can be seen that after removal the 
problem of excess AM is eliminated and the profiles 
show a drop at low values of $s$, suggesting $\alpha>1$. 
These profiles are qualitatively similar to the
ones expected for exponential discs
\citep[see][]{2005ApJ...628...21S,2001MNRAS.326.1205V}. 
Although, high AM material is not removed but the curve at high 
AM end is still shifted to the left. This is because 
removing  the low AM material increases the  $j_{\rm tot}$ of the system.
In \fig{removal_effect}, we
quantitatively assess the effect of removing the low AM material. 
We plot for various different values of $s_0$,  
the mean fraction of mass removed $f_{\rm reject}$ and the 
fraction of halos that can form exponential discs, i.e., halos with 
shape parameter $\alpha>1.3$. 
The higher the value of $s_0$,
the greater the amount of material removed 
and higher the fraction $f(\alpha>1.3)$. The figure shows 
that if $90\%$ of halos are to host exponential discs then one requires 
$f_{\rm reject}$ to be $0.4$.

\begin{figure}
   \centering \includegraphics[width=0.5\textwidth]{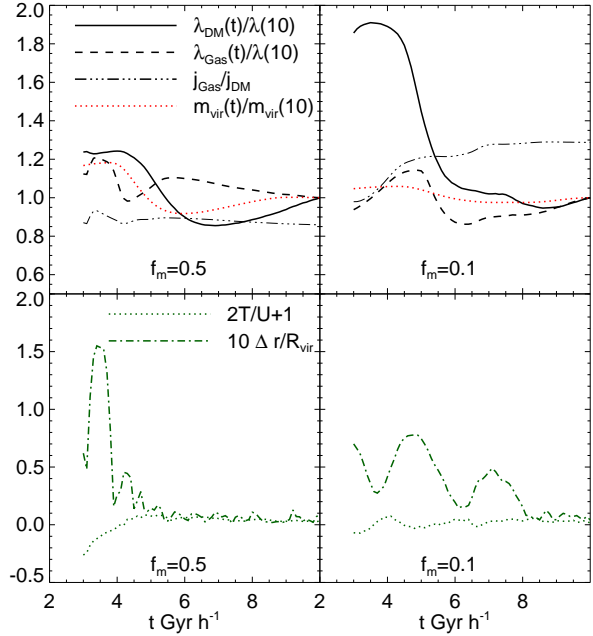}
   \caption{Evolution of spin, total angular momentum, virial mass, offset
     parameter and the virialization parameter with time. Results are
     shown for mass ratio $f_m=0.5$ (left panels) and $f_m=0.1$ (right
     panels). $\lambda(10)$ and $m_{\rm
     vir}(10)$ refer to values of spin and virial mass at $t=10 \gyr
   \hinv$ and are used to normalize the values of spin and mass with respect to the
   final equilibrium distribution. 
\label{fig:am_evolution}}
\end{figure}
\subsection{Spin Up and Spin Down of Halos Accompanied by Mergers}
It has been reported in earlier studies that 
immediately after the merger, e.g., the point of pericentric passage,
the spin parameter of the dark matter halo is found to be higher and 
later on as the system virializes the spin is found to drop. 
To study this we plot in
\fig{am_evolution}  the evolution of spin parameter of gas and dark
matter for two merging scenarios, mass fraction $f_m=0.5$ (Sim-1) and
$f_m=0.1$ (Sim-7). 
Note, for other merging scenarios the results are quite similar to the 
Sim-1 case.
In both panels $\lambda_{\rm DM}$ is found to drop sharply from the point of
first pericentric passage with a slow subsequent rise later on. The 
evolution of $\lambda_{\rm Gas}$ is sensitive to the choice of $f_m$ 
but in general shows much less variation than that of $\lambda_{\rm DM}$. 
The drop  in $\lambda_{\rm DM}$ for $f_m=0.5$ case is about  $30\%$
whereas for $f_m=0.1$ is about $80\%$. 
The sharpest fall of $\lambda_{\rm DM}$ is found to last for about $1 \hinv\gyr$ 
and occurs between $t=4 \hinv\gyr$ to $t=5 \hinv\gyr$ in the simulations. 

Next, we look at the evolution of the ratio 
${j_{\rm Gas}/j_{\rm DM}}$.
For $f_m=0.5$, the gas has lost some AM 
by the time of the start of the merger, but after that the ratio 
${j_{\rm Gas}/j_{\rm DM}}$ seems to remain constant. On the other hand,  
for $f_m=0.1$ case, the gas is found to gain AM from dark matter.  

Normally, the virialization ratio $2T/U+1$ and the 
offset parameter defined as $\Delta r/r_{\rm vir}$ is used to detect
such non relaxed halos. These are also plotted alongside. 
Here, $T$ is the kinetic energy and $U$ the potential
energy of the system and the offset is defined by  
$\Delta r=|{\bf x}_{\rm cm}-{\bf x}_{\rm max\ den}|$. The most commonly used values 
of these quantities are $-0.5<2T/U+1<0.5$ and $\Delta r/r_{\rm
  vir}<0.1$. It can be seen from \fig{am_evolution} that both these 
criteria have limited effect in detecting such cases. 
Our results suggest that a choice of $\Delta r/r_{\rm  vir}<0.025$
should be more effective in detecting such high spin systems.

\section{Discussion and Conclusions}
We have performed non-radiative hydrodynamical simulations of mergers 
of spherical
halos with a view to understand the AM properties  
of halos simulated in a cosmological context. 
The simulations being non-radiative
are fundamentally aimed at
determining the AMD of the gas before it cools onto the 
disk. In reality, at the time of most mergers  
a condensed component of gas also exists in addition to the hot 
gaseous halo. This cold component has not been taken into account in 
our simulations. This cold component plays a major role 
in fueling the growth of black holes and quasar activity. 
However, a significant fraction of this is also driven away due 
to feedback \citep{2005ApJ...630..705H,2006ApJS..163....1H}. 
The relative proportion of hot and cold component during a merger,  
their  interaction with each other and their 
relative contribution to the final disc is still not 
fully understood. 
In this light, our results concerning properties of disc galaxies are mostly  
applicable for discs (or its parts) that are predominantly formed out 
of gas accreted from the hot gaseous halo. 
The main AM properties studied include the evolution of
AM, the spatial distribution of AM and 
the orientation of the AM within the halo.
We also explored the differences between the AM properties 
of gas and dark matter and explain their origin. We now summarize our
results and discuss their implications for the formation 
of disk galaxies.

The shape parameter $\alpha$ of AMDs of gas in merger remnants is 
less than one for a wide variety of orbital parameters. 
This seems to be a generic result of the merging process.
Values greater than one and reaching upto 1.08 only occur for
unrealistically large
value of $\lambda$ or very low concentration parameter.
Lower values of mass ratio $f_m$ and higher values concentration 
parameter $c$ result in  lower value of $\alpha$ .
Under the assumption that disks form under conservation of AM 
this leads to
disks that are too centrally concentrated, as exponential disks 
require a value of $\alpha$ greater than $1.3$. In a previous study by  
\citet{2002MNRAS.335..487M} it was suggested that halos acquire most of their AM by means 
of  major mergers while minor mergers with small satellites, 
which come in from random directions, contribute to the low AM material. They argued that 
by preferentially discarding gas from  the shallow potential of these
small halos, e.g., by means of supernova feedback, the  AMD  
problem could be solved. However, our results  show  that even in absence of minor mergers, 
the AMD generated by a major merger has an excess of low AM 
material. Even the most favorable of merging scenario thus cannot account
for the formation of disk galaxies. Indeed, mergers in which the 
puffing up of gas by feedback was mimicked by decreasing the 
concentration parameter of the gas, did show a slight increase in value
of $\alpha$, suggesting it may partially help in reducing the low 
AM material but is not enough to solve the problem.

We find that the angular velocity $\Omega$ is almost independent of the spherical 
coordinate $\theta$ but exhibits a significant radial gradient. Such behavior is also seen 
for halos drawn from cosmological simulations. Hence, spherical shells
of matter appear to be moving in solid body rotation. This seems like
a safe assumption to be used for semi-analytic modelling \citep{2001MNRAS.327.1334V,2002MNRAS.332..456V}. 
The spatial distribution of low AM material is found to be in the
center and along a conical region along the rotation axis. This suggests that a mechanism which can preferentially eject
the material in the center and along the poles can 
alleviate the AMD problem. In fact, feedback from 
intense star formation in the inner regions can drive such an outflow
. Additionally, feedback from AGN jets is also expected to evacuate material along
the axis.
Evidence of such  conical outflows  is also provided by
observations
\citep{1990ApJS...74..833H,1998ApJ...493..129S,2002ApJ...565L..63V,2005ARA&A..43..769V,2007Ap&SS.311...87B}.
More recently, using oxygen absorption lines in background quasars, 
significant amounts of diffuse metal rich gas has been 
detected in halos of star forming galaxies, suggesting large scale 
star formation driven galaxy outflows
\citep{2011Sci...334..948T,2011Sci...334..952T}.
Essentially, during the formation of the
galaxy the star formation will be strongest in the central regions
and this will drive an outflow which will expand more along the rotation
axis due to the flattened geometry of the assembling gas. This will
prevent the rest of the low AM
material from falling onto the disc. This mechanism was initially discussed
in \citet{2005PhDT........20S} (Sec-3 and 5.2 of the PhD thesis), 
  and  more recently has also been proposed by
\citet{2011MNRAS.tmp..595B}. 
Using high resolution cosmological
simulations incorporating star formation and feedback
\citet{2011MNRAS.tmp..595B} clearly demonstrate that the above mechanism is responsible
for the formation of bulgeless dwarf galaxies \citep[see also][]{2011arXiv1103.6030G}. 
Additionally, as suggested by \citet{2011arXiv1105.2562B}, for small mass 
systems the outflows can eject the low
AM material but for large mass systems it can also drive a
fountain leading to mixing and redistribution of the low AM material 
to be accreted later on as high AM material.
In addition to cosmological simulations,
preferential ejection of low AM material by feedback 
from the central regions was also used in semi-analytic modelling 
by \citep{2009MNRAS.396..141D,2009MNRAS.396..121D} to successfully 
reproduce the exponential structure of discs. 

We tested a simple geometric criteria to selectively remove 
the low AM material from halos and found that resulting AMDs 
are in good agreement with those of exponential disc galaxies. 
Our results suggest that in order for $90\%$ of halos to form 
realistic exponential  discs approximately $40\%$ of the baryonic 
mass needs to be rejected. 
The presented criteria can be easily applied to dark matter simulations 
or in semi-analytical modelling. 
Note, our results provide an upper limit 
on the fraction of retained baryons
$f_{d}=1-f_{\rm reject}=M_{\rm disc}/((\Omega_{\rm b}/\Omega_{\rm m})M_{\rm
  vir})=0.6$, $M_{\rm disc}$ being mass of stars and gas in
the  galactic disc. Real galaxies can have even lower $f_{\rm d}$ 
depending upon other physical processes which regulate the star 
formation efficiency.  In the context of the missing baryon problem,  
the upper limit given by AMD argument is in good agreement with observations.
\citet{2010MNRAS.404.1111G} 
using abundance of galaxies from SDSS find a maximum value of 
$f_{\rm d}$ to be $0.2$ at $M_{\rm vir}=6\times10^{11}$. 
\citet{2010ApJ...708L..14M}  estimate $f_{\rm d}<0.4$ for spiral
galaxies and even lower for dwarfs. Note, \citet{2010ApJ...708L..14M} 
use an overdensity factor of $\Delta=500$ (instead of $100$) to 
measure the mass of the halo, 
so actual values of $f_{\rm d}$ with in virial radius 
would be about a factor of two lower.

The difference between collisionless dynamics and gas 
dynamics results in differences between the AM
properties of the gas and dark matter and this can
potentially have implications for studies that assume them to be same.
The gas as compared to dark matter is more efficient in depositing 
its orbital AM in the central parts of the halo. 
This results in a higher value of the spin parameter 
$\lambda$ for the gas as compared to dark matter and also a moderately high value 
of the shape parameter $\alpha$. Lower values of mass fraction $f_m$
and initial concentration also result in higher 
$\lambda_{\rm Gas}/\lambda_{\rm DM}$
while lower values of $\lambda_{\rm initial}$ and merging time $t_{\rm orb}$
result in lower values of spin ratio. 
About $6 \hinv\gyr$ after the merger, i.e., the first pericentric
passage, the ratio  $\lambda_{\rm
  gas}/\lambda_{\rm DM}$ is found to be greater than 1 for all merging 
scenarios analyzed here. This seems to be consistent with spin ratios of halos 
obtained from cosmological simulations at $z=0$, where $<\lambda_{\rm
  gas}/\lambda_{\rm DM}>$ is close to 1.4. Using a 
sample of 14 dwarf galaxies \cite{2001MNRAS.326.1205V} had found the
median spin of galaxies to be $0.06$, assuming $\lambda_{\rm DM}=0.0367$
this gives $\lambda_{\rm  gal}/\lambda_{\rm DM}=1.63$. 
The higher spin of galaxies could be due to the gas having 
higher spin, but it could also be due to preferential rejection of low AM 
material during the assembly of the disk.

We find that for mergers with zero intrinsic spins, the AM vectors of 
gas and dark matter are well aligned with misalignment angle being less
than $2^{\circ}$.  On the other hand, mergers having non-zero  
intrinsic spins which are inclined at an angle to the orbital 
AM vector can result in a misalignment of about $20^{\circ}$, 
consistent with halos simulated in a cosmological context.
Since halos simulated in cosmological context undergo multiple
mergers with different spin orientations, the above result 
provides a natural explanation for this. 
This shows that the misalignment can be explained purely 
by means of mergers without any need for the gas and the dark matter 
to be torqued differently during the formation of the proto-halo.
In general, the gas within the virial radius 
is more effective in retaining the information about the intrinsic
spins of the merging halos whereas the dark matter is more effective 
in retaining the orbital AM information. The misalignment between
gas and dark matter has important implications for studies such as,  
the correlation between the anisotropic distribution of satellite
galaxies and the major axis of the central galaxy and weak lensing
studies attempting to measure the ellipticity of the 
dark matter halos.

Mergers with non aligned spins also tend to make the 
AM of gas as measured in radial shells 
misaligned with each other.
Since galaxies generally form inside-out, later infall of 
misaligned material can cause warps in disk galaxies,  
and this has recently been shown by \citet{2010MNRAS.408..783R} 
in cosmological hydrodynamical simulations with star formation. 
They find that immediately after the major merger the 
inner gas which forms the disk 
is misaligned with the rest of the gas in the halo. 
Later infall 
of misaligned gas causes the warps. 
A probable explanation for the cause of misalignment is given by 
them as the fact that interactions such as minor mergers can affect 
AM of the inner and outer regions differently. 
We have here explicitly
demonstrated as to how major mergers, in which the orbital AM 
is not aligned with the intrinsic spin of the halos,  
generates such a misalignment. Minor mergers later on 
may further alter the orientation but are not necessarily required
to generate the misalignment. 

Our results show that the orientation of the 
AM 
within the halo depends sensitively upon the orientation of the
intrinsic spins of the merging halos with respect to that of the
orbital AM. The larger the initial misalignment between
the initial AM vectors the larger the final misalignment 
between the inner and outer parts. This suggests that warps may 
offer the possibility to probe the merger history of the halo.

Observational evidence for warps is quite ubiquitous
\citep{1976A&A....53..159S,1990ApJ...352...15B,1994AJ....108..456R,2002A&A...389..825V}. \citet{2002A&A...394..769G} find that in their sample of galaxies, all galaxies that have an extended HI disk with respect to the optical are warped. 
If misalignments in AM are not as frequent then this
could pose a problem. Our results show \figp{misalignment_sim} 
that AM of 
gas within $r<0.1r_{\rm  vir}$ and $r>0.9r_{\rm vir}$ is misaligned by
more than $10^{\circ}$, with  respect to the total AM
vector,  for about $84\%$ of the halos (the median misalignments being
about $30^{\circ}$ and $17^{\circ}$ respectively). 
This demonstrates that the
misalignments are quite common and supports the idea that they 
are responsible for warps. Just as perfect prograde and perfect
retrograde mergers are rare so are systems with small angle warps 
and systems with counter rotating gas. In future, observations with 
detailed statistics on the orientation of the warps could be employed 
to check if they match with the distribution of misalignments
predicted by theory.

As mentioned earlier, the amount of misalignment in general is found 
to increase with the increase of 
angle between the orbital and intrinsic AM vectors. 
For retrograde encounters the gas at intermediate radii is even found to 
be counter rotating. This could be responsible for the counter rotating
gas seen in some galaxies.  
Generally, mergers of gas rich systems are invoked to explain such
systems. The quantity of counter rotating gas in some galaxies 
such as NGC3626 is so large that a single minor merger cannot properly 
account for it. If on the other hand a merger is not minor then it 
can heat up and thicken the disk considerably. 
A slow, continuous 
and well dispersed accretion, as opposed to an accretion via  
a merging system is preferred
\citep{1995Natur.375..661C,1996ApJ...461...55T}. 
Counter rotating gas in galactic 
halos formed by retrograde mergers as shown here, naturally 
provides such an extended reservoir of gas. 
A recent merger which can potentially heat up the disk 
is not required, the counter rotating gas is formed early on 
during the last major merger, which causes the
inner and outer regions to rotate in different directions.  
In such a scenario, the inner regions first assemble to form 
the disc, rest of the material falls later on to generate 
the counter rotating gas.

We also studied the issue of spin up of a halo undergoing a merger 
and the subsequent spin down during virialization \citep{2001ApJ...557..616G,2002ApJ...581..799V,2004MNRAS.348..921P,2006MNRAS.370.1905H}. 
As argued by \citet{2007MNRAS.380L..58D}, in collisionless mergers 
central regions tend to be populated by low AM material 
and high AM material is pushed to weakly bound orbits. 
When AM is measured with the fixed radius, such as  
the virial radius, the effect is a spin down.   Our merger simulations 
also show a similar effect. For dark matter the inner half loses 
AM while the outer half gains.
The main cause for such a redistribution of AM is 
the collisionless dynamics and is as follows. 
For dark matter during the collision the late in-falling particles 
have high AM and high energy and they gain energy 
during the collapse and hence can easily climb out of the final relaxed
potential which is much shallower. Hence, high AM
particles get pushed to weaker and weaker orbits 
making them move outwards.
For gas the late in-falling particles shock and deposit their 
AM onto the inner regions making them behave differently. 
We find that 
for a Milky Way sized halo the spin down process lasts about a giga year, 
and the spin of dark matter can fall by about $40-80\%$ during this time, depending
mainly upon the mass ratio of the merging components.
The spin of gas shows somewhat less 
variation.  We find that  
the virial ratio $2T/U+1$ is not very effective in detecting such 
situations. The offset parameter is more successful in detecting 
such cases but a value of $\Delta r/r_{\rm vir}<0.025$ would be
needed, which is much less than what is currently used
\citep{2007MNRAS.380L..58D,2007MNRAS.381.1450N}.
 Hence, recent results showing high spin 
systems to be more clustered may be affected by this bias 
\citep{2007MNRAS.376..215B,2010MNRAS.407..691D}. 
Alternatively, it may reflect the fact that in clustered
environments mergers and hence non relaxed halos are more common. 
If non relaxed halos is the cause of correlation between the 
clustering and spin 
then the ability to observationally detect it by measuring spin of
galaxies is unclear,  as galaxies form out of baryons in relaxed halos. 
Additionally, it  is not known if the spin of baryons would also 
show such a clustering.

Finally, our results show that mergers of NFW halos naturally generate 
the universal form of AMDs
as seen in simulations. For dark matter the
value of the shape parameter $\alpha$ is in excellent agreement with 
the results from cosmological simulations. However, this does not mean 
that mergers are the only way to generate such distributions. As has 
been shown recently by \citet{2009MNRAS.396..709W} even hot dark matter 
simulations which have almost no mergers show such AMDs. Hence, the origin 
of the universal form is more generally related to the virialization 
processes such as the violent relaxation. 
However, our results show 
that mergers do induce subtle differences between the AM
properties of dark matter and gas. 
The alignment of the AM
vector within the halo and also that of gas with respect to
dark matter is sensitively related to the merger 
history and may serve to discriminate the dark matter models.  
In hot dark matter models, although less likely, 
misalignments as discussed above could also be produced if 
matter coming from different regions have AM pointing in different
directions.

\section*{Acknowledgments}
We would like to thank the anonymous referee for the comments and
suggestions. This work has been supported by grants from the U.S. 
National Aeronautics
and Space Administration (NAG 5-10827), the David and Lucile Packard
Foundation. SS is funded through ARC DP grant 0988751
which supports the HERMES project.
JBH is funded through a Federation Fellowship from the Australian 
Research Council (ARC). A part of the research was undertaken on the 
NCI National Facility in Canberra, Australia, which is supported by 
the Australian Commonwealth Government.

\appendix
\section{Orbital Parameters} \label{sec:merger_param}
\begin{figure}
   \centering \includegraphics[width=0.45\textwidth]{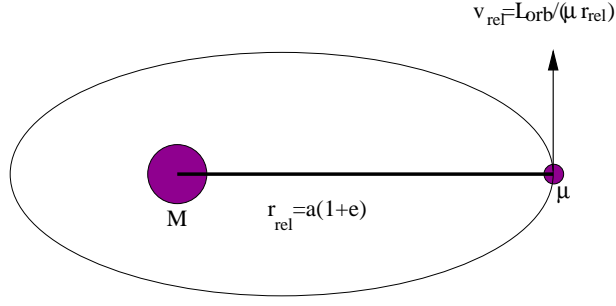}
   \caption[Orbital Parameters]{ Merger of two halos can be reduced to a one body problem of
     mass $\mu$ moving in the potential of mass $M$. The orbit can be
     characterized by semi-major axis $a$ and eccentricity $e$. At maximum 
     separation $r_{\rm rel}=a(1+e)$ the tangential velocity $v_{\rm rel}$ is given 
     by the angular momentum acquired by the masses during the expansion phase. 
\label{fig:merger}}
\end{figure}

The merger of two bodies of mass $m_1$ and $m_2$ can be reduced to the motion
of a test particle, with a reduced mass $\mu  =m_1m_2/(m_1+m_2) $, 
in the potential of a mass $M=m_1 + m_2$ (\fig{merger}). The initial conditions are set by specifying the
relative separation ${\bf r}_{\rm rel}$ and relative velocity ${\bf v}_{\rm rel}$.
In a cosmological context the 
two masses first move apart  due to Hubble expansion and eventually, come to a
halt and collapse due to their mutual gravitational attraction. The orbits of interest 
are those which are bound and collide within a Hubble time.
A bound orbit can be fully
characterized by its eccentricity $e$ and the semi-major axis $a$.
The energy of the orbit $E_{\rm orb}$, and the orbital time period $T_{\rm orb}$ are related
to $a$ by
\be
E_{\rm orb} & =& -\frac{GM\mu}{2a} \textrm{    \ \ \ \ ,\ \ \ \      } T_{\rm orb}  =  2\pi\sqrt{\frac{a^3}{GM}} 
\ee
$E_{\rm orb}$ can be written in terms of $T_{\rm orb}$ as
\be
E_{\rm orb} & =& -\frac{1}{2}(4\pi^2G^2)^{1/3} T^{-2/3}_{\rm orb} f_{\mu} M^{5/3} 
\ee
where 
$f_{\mu} = \frac{\mu}{M}$.
The angular momentum $L_{\rm orb}$ is related to eccentricity $e$ by
\be
L_{\rm orb} & =& \mu \sqrt{GMa} \sqrt{1-e^2} = \frac{GM^{5/2}f_\mu^{3/2} \sqrt{1-e^2}}{\sqrt{2|E_{\rm orb}|}} 
\ee

According to the tidal torque theory the system acquires AM 
during its expansion phase, with the AM  increasing nearly
linearly  with time  during the initial linear phase
of growth of density perturbations \citep{1984ApJ...286...38W,1970Afz.....6..581D}. 
The acquisition of AM ceases in the non-linear regime. 
We assume that all AM is acquired by the time  of maximum expansion which gives
\be
r_{\rm rel} & = &a(1+e) 
\ee
At maximum expansion, the radial velocity being zero, the total velocity is  
given by the tangential velocity.
\be
v_{\rm rel} & = & \frac{L_{\rm orb}}{\mu r_{\rm rel}} 
\ee
Since the merging bodies are extended objects, the total energy is given by the sum of the orbital energy plus the self energy of the bodies.
The self energy of a body of mass $M_v$ and radius $R_v$, having an NFW density
profile \citep{1996ApJ...462..563N,1997ApJ...490..493N} with concentration parameter $c$, is given by \footnote{For the Einasto profile the formulas are
  available at \citet{2009ApJ...707.1642N,2011ApJ...732...17N}}
\be
\label{equ:ev}
E_v & = & -f_c\frac{GM_v^2}{2R_v} \textrm{\ \ \ \ where\ \ } f_c  =  \frac{c}{2}\frac{1-1/(1+c)^2-2\textrm{ln}(1+c)/(1+c)}{(\textrm{ln}(1+c)-c/(1+c))^2} 
\ee

Assuming that both halos are virialized at a redshift of $z$,  
$M_v$ can be written in terms of $R_v$ are as
\be
M_v  &= & \frac{4\pi R_v^3}{3} \Delta(z) \frac{3H^2(z)}{8\pi G} =   R_v^3 \frac{\Delta(z) H^2(z)}{2G},
\ee
where $\Delta(z)$ is the over-density criteria used to identify a 
virialized region, i.e., a spherical region whose average mass density is $\Delta(z)$
times the critical density at that redshift. 
$\Delta(z)$ is
approximated by \citep[]{1998ApJ...495...80B} $ \Delta(z) \simeq
(18\pi^2 +82x+-39x^2)$, where $x=\Omega_{\rm m}(z)-1$.
Consequently, the total energy is given by $E  = E_{v1} +E_{v2}+E_{\rm orb}$.
Analogously, the spin parameter $\lambda$ of the whole system is given by
\be
\lambda &= & \frac{L|E|^{1/2}}{GM^{5/2}}  = \sqrt{\frac{1-e^2}{2}}f_{\mu}^{1.5}\left(  \frac{|E|}{|E_{\rm
      orb}|}\right)^{1/2} 
\ee
Instead of the semi-major axis $a$ and the eccentricity $e$
the orbit can  be equivalently parameterized in terms of  
the orbital time period $T_{\rm orb}$ and the spin parameter
$\lambda$. 
We restrict ourselves to values of $T_{\rm orb}$ which have $r_{\rm rel}>r_{12}$ where $r_{12}=r_{\rm vir1}+r_{\rm vir2}$.

\section{Mass Structure Of Remnant Halos} \label{sec:mass_struct}
\begin{figure}
   \centering \includegraphics[width=0.45\textwidth]{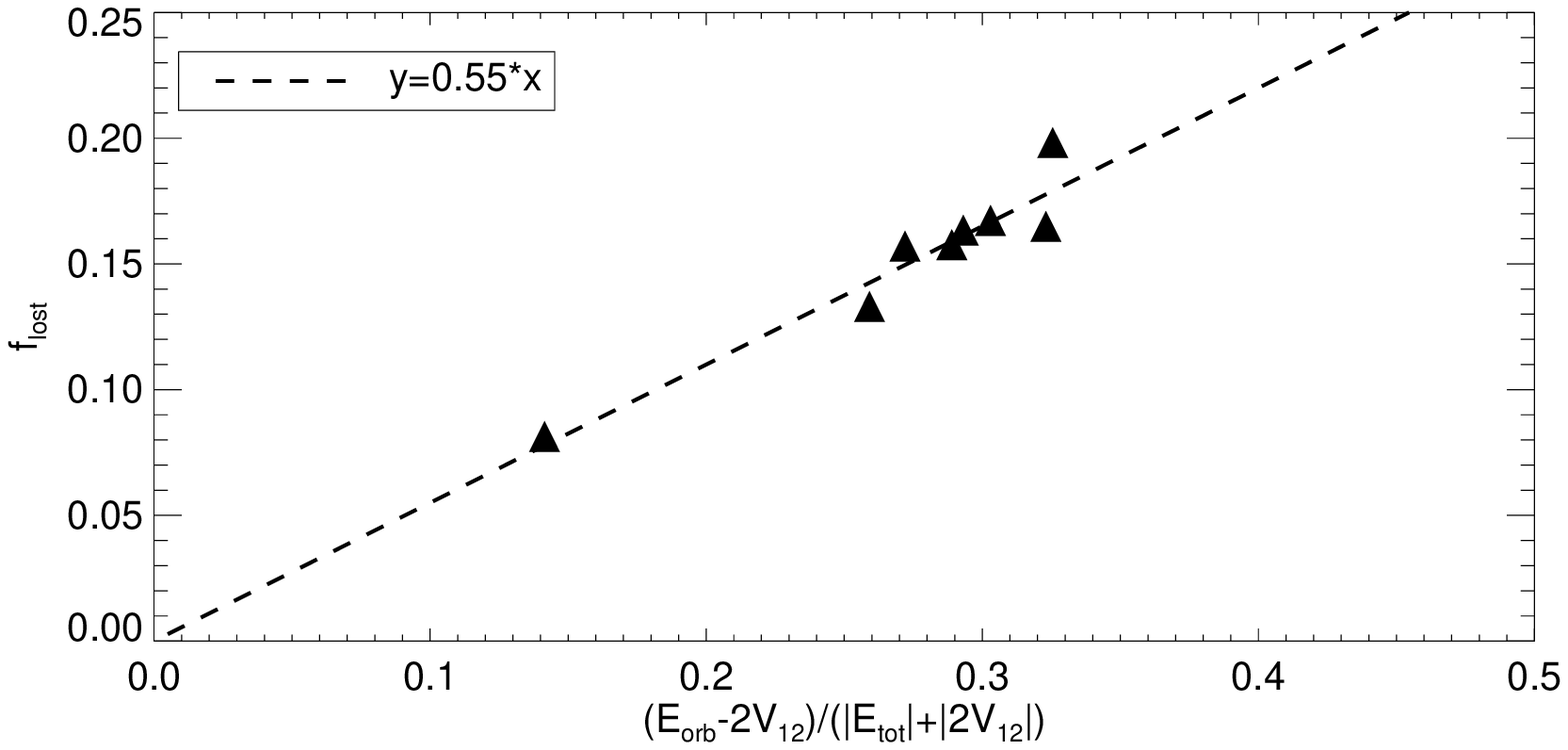}
   \caption[ Dependence of fraction of mass lost during a collision on merger 
     parameters.]{ Dependence of fraction of mass lost during a collision on merger 
     parameters. The fraction of mass lost is an increasing function of the
     kinetic energy involved in the collision and a decreasing function of the
     total binding energy of the system.
 \label{fig:flost_e}}
\end{figure}

The final properties of the merger remnants are given in \tab{tb1}.
We note that the virial mass $m_{\rm vir}$ of the remnant is less than the total
mass of the system $m_{\rm tot}$. Hence, a fraction of mass is lost
which we define as $f_{\rm lost}=(m_{\rm tot}-m_{\rm vir})/m_{\rm tot}$. 
Also, the concentration parameter of the remnant halo
$c_{\rm final}$ is slightly larger than $c_{\rm initial}$.

It is interesting to know if the final properties of the halo e.g., $m_{\rm vir}$
and $c_{\rm final}$ can be predicted from the initial conditions.
We expect  the fraction of lost mass 
$f_{\rm lost}$ to be an increasing function 
of the kinetic energy $KE$ involved in the collision and a decreasing function of the 
total binding energy of the system. We find that the following
empirical formula, which satisfies the above conditions, fits the results 
obtained from simulations (\fig{flost_e}).
\be
f_{\rm lost} & \propto &  \frac{\textrm{Maximum KE of collision at
    $r_{\rm sep}=r_{12}/2$ }}{\textrm{$|E_{\rm tot}|$ + PE at $r_{\rm sep}=r_{12}/2$}} \\
&= & k_{f} \frac{E_{\rm orb}-2V_{12}}{|E_{\rm tot}|+|2V_{12}|} \textrm{\ \ \ \ where \ } V_{12}=-\frac{GM\mu}{r_{12}} 
\label{equ:f_lost}
\ee
If $c_{\rm initial}$ is higher the system has higher $|E_{\rm tot}|$ consequently, it is more
bound and loses less mass. If $|E_{\rm orb}|$ is higher the system is again more
bound and also the KE of the collision is less, consequently  reducing the mass loss.

Interestingly, the total energy of the remnant halo $E_{\rm vir}$
(putting $c_{\rm final}$ and $m_{\rm vir}$ from \tab{tb1} in
\equ{ev})
is nearly equal to the energy of the system $E_{\rm total}$ before the
merger. This suggests that the mass that lies outside the virial 
radius, consists of a bound and an unbound part and has almost zero net energy. Consequently 
the concentration parameter of a remnant halo can be predicted from 
the knowledge of its orbital parameters.

\bibliographystyle{apj} 
\bibliography{mybib}

\end{document}